%% file: 0_main.tex
\documentclass[lettersize,journal]{IEEEtran}

\usepackage{amsmath,amsfonts}
\usepackage{algorithmic}
\usepackage{algorithm}
\usepackage[algo2e,ruled,vlined]{algorithm2e}
\usepackage{array}
\usepackage{textcomp}
\usepackage{stfloats}
\usepackage{url}
\usepackage{verbatim}
\usepackage{graphicx}
\usepackage{cite}
\usepackage{soul}
\usepackage{multirow}
\usepackage{nicematrix}
\usepackage{xcolor}
\usepackage{colortbl}
\usepackage{caption}
\usepackage{subcaption}

\usepackage[colorlinks=true,linkcolor=blue,citecolor=blue,urlcolor=blue]{hyperref}

\begin{document}


\title{How Roadside Units Enhance Intersection Safety? Cooperative Autonomous Driving System Design and A Proof of Concept}

\author{
    Taoyuan Yu,~\IEEEmembership{Student Member,~IEEE,}
    Kui Wang,~\IEEEmembership{Student Member,~IEEE,}
    Zongdian Li,~\IEEEmembership{Member,~IEEE,}
    Tao Yu,~\IEEEmembership{Member,~IEEE,}
    Walid Saad,~\IEEEmembership{Fellow,~IEEE}
    and Kei Sakaguchi,~\IEEEmembership{Senior Member,~IEEE,}%
    
    \thanks{This work was supported by MONET Technologies Inc. (Responsible person: Toshiki Demizu), the Japan Science and Technology Agency (JST) SPRING JPMJSP2180, the Science Tokyo Academy for Super Smart Society, the U.S. National Science Foundation under Grant CNS-2210254, and the Japan National Institute of Information and Communications Technology (NICT) under JUNO Grant 22404.}%
    
    \thanks{T.~Yu, K.~Wang, Z.~Li, T.~Yu, and K.~Sakaguchi are with the Department of Electrical and Electronic Engineering, School of Engineering, Institute of Science Tokyo (Email: yuty,kuiw,lizd,yutao,sakaguchi@mobile.ee.titech.ac.jp).}%
    
    \thanks{W. Saad is with the Bradley Department of Electrical and Computer Engineering, Virginia Tech, Arlington, VA 22203, USA (Email: walids@vt.edu).}%
}

\maketitle
\thispagestyle{plain} 

\begin{abstract}

Intersections remain one of the most hazardous locations in urban road networks, where heterogeneous traffic participants and limited visibility frequently lead to severe traffic conflicts. In this paper, a vehicle-to-infrastructure-to-vehicle (V2I2V) cooperative system is proposed for improving road safety and traffic efficiency by using digital twins (DTs) deployed on roadside units (RSUs) to eliminate blind spots and centrally coordinate connected and automated vehicles (CAVs) in smart intersections.  The proposed system integrates cloud-based global DTs for macroscopic guidance and RSU-based local DTs for real-time operations. Within this architecture, a hierarchical reinforcement learning (HRL) framework combines offline pre-training with online fine-tuning to achieve robust cooperative control. Experimental results show that the proposed system achieves substantial improvements in safety and efficiency in simulation experiments and real-world proof-of-concept (PoC) trials. In simulations, our system ensures high safety, efficiency, and smoothness under realistic communications and traffic constraints. In PoC trials, the RSU-centric control loop achieves a decision-making latency of approximately 42\,ms and maintains a safe stopping distance of 8.5\,m for pedestrians, while also shortening stop duration and overall traversal time. These results indicate that the proposed system provides robust and scalable performance at smart intersections.
\end{abstract}


\section{Introduction}
\input{1_intro}

\section{DT-based V2I2V Cooperative System}
\input{2_SMDT}

\section{Cooperative Driving Algorithm Design}
\input{4_implementation}

\section{Simulation and Comprehensive Performance Evaluation}
\input{5_Sim}

\section{Experimental Platform and Environment Setup}
\input{6_PoCset}

\section{Proof-of-Concept Results}
\input{7_evaluation}

\section{Conclusion}
\input{8_conclusion}

\bibliographystyle{IEEEtran}
\bibliography{bibliography}

\begin{IEEEbiography}
[{\includegraphics[width=1in,height=1.25in,clip,keepaspectratio]{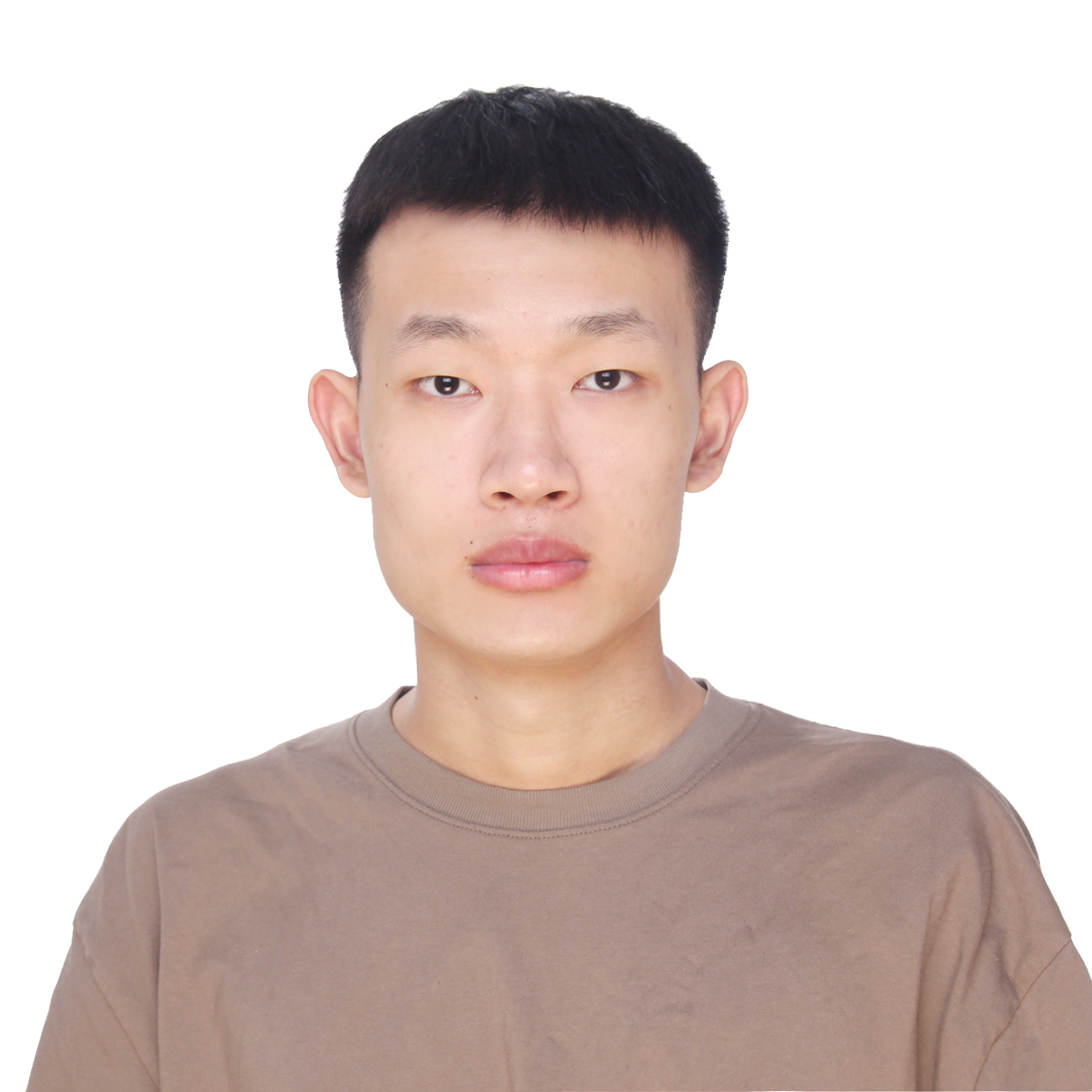}}]{Taoyuan Yu} (Student Member, IEEE)  received the B.E. degree in electronic information engineering from Southwest University, Chongqing, China, in 2022, and the M.S. degree in electrical engineering from Washington University in St. Louis, St. Louis, USA, in 2023. He is currently working toward the Ph.D. degree in the Department of Electrical and Electronic Engineering, Institute of Science Tokyo, Tokyo, Japan. His current research interests include autonomous driving, V2X communication, deep learning, reinforcement learning.
\end{IEEEbiography}

\begin{IEEEbiography}
[{\includegraphics[width=1in,height=1.25in,clip,keepaspectratio]{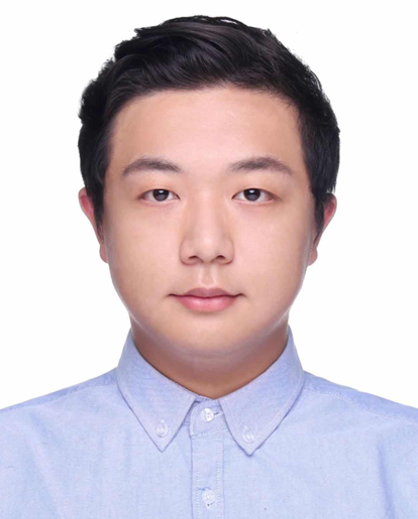}}]{Kui Wang} (Member, IEEE) received the B.E. degree in mechanical engineering from the Beijing University of Posts and Telecommunications, China, in 2018, the M.E. degree in vehicle engineering from the KTH Royal Institute of Technology, Sweden, in 2021, and the Ph.D. degree in electrical and electronics engineering from Institute of Science Tokyo, Japan, in 2025. He is a postdoctoral researcher with the Department of Electrical and Electronic Engineering, Institute of Science Tokyo. His research interests include digital twins, smart ocean, autonomous driving, and machine learning.
\end{IEEEbiography}

\begin{IEEEbiography}[{\includegraphics[width=1in,height=1.25in,clip,keepaspectratio]{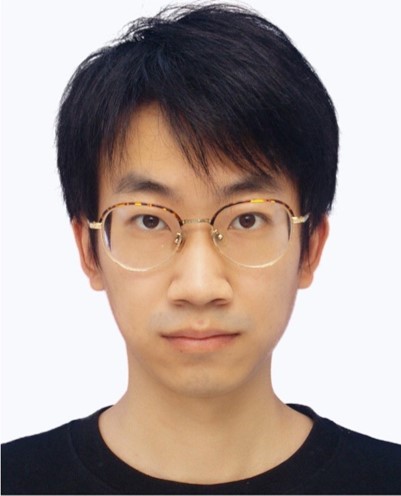}}]{Zongdian Li} (Member, IEEE) received the B.E. degree in communication engineering from Beijing University of Posts and Telecommunications, Beijing, China, in 2018, and the M.E. and Ph.D. degrees in electrical and electronic engineering from Tokyo Institute of Technology, Tokyo, Japan, in 2020 and 2023, respectively. Currently, he is an Assistant Professor with the School of Engineering, Institute of Science Tokyo (formerly Tokyo Institute of Technology). His research interests include vehicle-to-everything (V2X) communications, smart mobility, and digital twins. He is the members of IEEE and IEICE.
\end{IEEEbiography}

\begin{IEEEbiography}[{\includegraphics[width=1in,height=1.25in,clip,keepaspectratio]{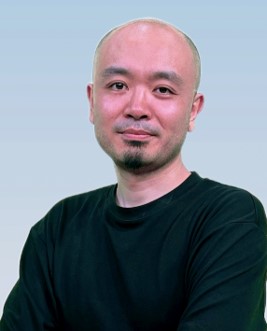}}]{Tao Yu} (Member, IEEE)received a M.E. degree from the Communication University of China in 2010, and a Dr.Eng. degree from the Tokyo Institute of Technology (now Institute of Science Tokyo, Science Tokyo) in 2017. After completing doctorate, he worked as a researcher from 2017 to 2022 in Dept. Electr. Electron. Eng. at Tokyo Tech. Since 2022, he has been a specially appointed associate professor at the Academy for Super Smart Society (now Academy of Super Smart Society, Science Tokyo). His research interests include smart mobility, autonomous driving, digital twin, V2X, UAV communication, mmWave, sensor networks, localization, antenna design, and building energy management. He is a member of IEEE and IEICE.
\end{IEEEbiography}

\begin{IEEEbiography}[{\includegraphics[width=1in,height=1.25in,clip,keepaspectratio]{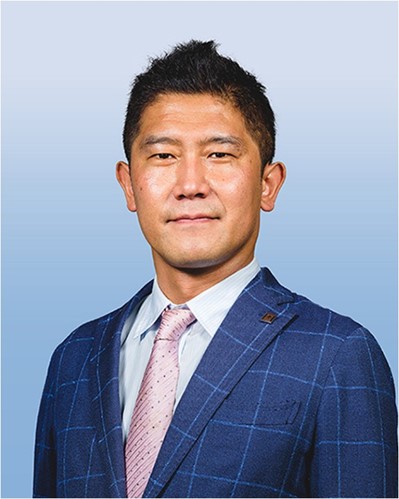}}]{Kei Sakaguchi} (Senior Member, IEEE) received the M.E. degree in information processing from Tokyo Institute Technology, Tokyo, Japan, in 1998, and the Ph.D. degree in electrical and electronics engineering from Tokyo Institute Technology, Tokyo, Japan, in 2006. He is currently working with the Institute of Science Tokyo in Japan, as the Dean with Science Tokyo Academy for Super Smart Society and as a Professor with the School of Engineering. His current research interests include 5G/6G cellular networks, millimeter-wave communications, wireless energy transmission, V2X for automated driving, and super smart society. He was the recipient of the Outstanding Paper Awards from SDR Forum and IEICE, in 2004 and 2005, respectively, and three Best Paper Awards from IEICE communication society in 2012, 2013, and 2015. He was also the recipient of the Tutorial Paper Award from IEICE communication society in 2006. He is a fellow of IEICE.
\end{IEEEbiography}

\begin{IEEEbiography}[{\includegraphics[width=1in,height=1.25in,clip,keepaspectratio]{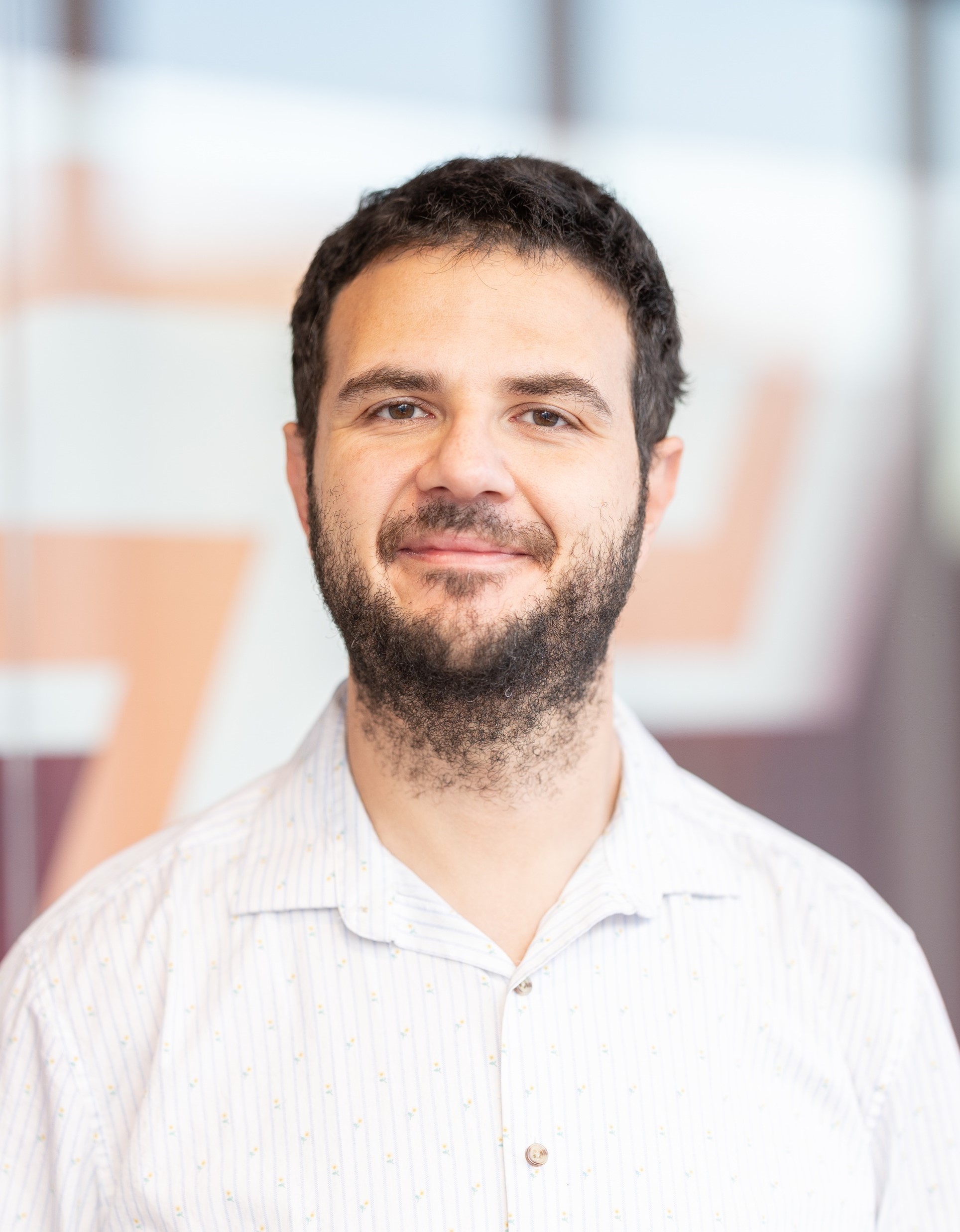}}]{Walid Saad} (Fellow, IEEE) received his Ph.D degree from the University of Oslo, Norway in 2010. He is currently a Professor at the Department of Electrical and Computer Engineering at Virginia Tech, where he leads the Network sciEnce, Wireless, and Security (NEWS) laboratory. His research interests include wireless networks (5G/6G/beyond), machine learning, game theory, security, UAVs, semantic communications, cyber-physical systems, and network science. Dr. Saad is a Fellow of the IEEE. He is also the recipient of the NSF CAREER award in 2013, the AFOSR summer faculty fellowship in 2014, and the Young Investigator Award from the Office of Naval Research (ONR) in 2015. He was the (co-)author of twelve conference best paper awards at IEEE WiOpt in 2009, ICIMP in 2010, IEEE WCNC in 2012, IEEE PIMRC in 2015, IEEE SmartGridComm in 2015, EuCNC in 2017, IEEE GLOBECOM (2018 and 2020), IFIP NTMS in 2019, IEEE ICC (2020 and 2022), and IEEE QCE in 2023. He is the recipient of the 2015 and 2022 Fred W. Ellersick Prize from the IEEE Communications Society, of the IEEE Communications Society Marconi Prize Award in 2023, and of the IEEE Communications Society Award for Advances in Communication in 2023. He was also a co-author of the papers that received the IEEE Communications Society Young Author Best Paper award in 2019, 2021, and 2023. Other recognitions include the 2017 IEEE ComSoc Best Young Professional in Academia award, the 2018 IEEE ComSoc Radio Communications Committee Early Achievement Award, and the 2019 IEEE ComSoc Communication Theory Technical Committee Early Achievement Award. From 2015-2017, Dr. Saad was named the Stephen O. Lane Junior Faculty Fellow at Virginia Tech and, in 2017, he was named College of Engineering Faculty Fellow. He received the Dean's award for Research Excellence from Virginia Tech in 2019. He was also an IEEE Distinguished Lecturer in 2019-2020.  He has been annually listed in the Clarivate Web of Science Highly Cited Researcher List since 2019. He currently serves as an Area Editor for the IEEE Transactions on Network Science and Engineering and the IEEE Transactions on Communications. He is the Editor-in-Chief for the IEEE Transactions on Machine Learning in Communications and Networking.
\end{IEEEbiography}

\vfill

\end{document}

%% file: 1_intro.tex
\subsection{Background}
\IEEEPARstart{E}{very} year, approximately 1.19 million lives are prematurely lost due to road traffic crashes\cite{who2025roadtraffic}. Among different traffic scenarios, intersections are particularly complex, as diverse road users compete for limited spatio-temporal resources, often leading to severe conflicts\cite{10976401}. This challenge is particularly critical in Japan, where narrow roads and densely residential areas result in short or acute-angle intersections that restrict drivers’ visibility, creating blind spots and increasing accident risks\cite{tian2025impactbuildinginducedvisibilityrestrictions}. According to the 2024 report by Asahi Shimbun, the rate of fatalities and injuries among pedestrians and bicyclists on roads narrower than 5.5 meters was 1.8 times higher than that on wider roads\cite{yoshida2024speedlimit}. These findings indicate that enhancing traffic safety at intersections has become an urgent priority for urban transportation systems\cite{yu2025digitaltwinbasedcooperativeautonomous}.

With recent advances in intelligent transportation systems (ITS), connected and automated vehicles (CAVs) can communicate with surrounding smart entities, such as vehicles and roadside units (RSUs) \cite{10976401}. By using vehicle-to-vehicle (V2V) and vehicle-to-infrastructure (V2I) communications, CAVs can broaden their sensing range beyond onboard sensors, and enhance their decision-making with shared information, enabling a more comprehensive understanding of the surrounding environment \cite{10729655, DONG2024107338}. This communications paradigm provides the foundation for vehicle-infrastructure systems, which leverage cooperative perception and centralized decision-making to improve safety and efficiency at smart intersections \cite{10623524}.

Within the vehicle-infrastructure systems, the integration of digital twin (DT) technology provides new opportunities for improving safety at smart intersections\cite{10964767}. DT creates a real-time virtual replica of the traffic environment\cite{10443037}, enabling interactive analysis of traffic dynamics and evaluation of control strategies. By offering a more comprehensive and reliable representation of the surrounding environment, DT supports earlier hazard anticipation, reduces blind-spot risks, and enhances decision-making at smart intersections\cite{11059924}.

Although significant progress has been made, existing studies still provide incomplete solutions for enhancing safety and coordination at smart intersections \cite{1011458}. Prior works have explored improvements through improved perception, enhanced vehicle coordination, and strengthened infrastructure support. To provide a foundation for our proposed system, the following section reviews approaches related to DTs, vehicle–infrastructure systems, and decision-making algorithms.

\subsection{Related Works}

\begin{table*}[ht]
\centering
\caption{Feature comparison of representative related works.}
\label{tab:feature_compare}
\begin{tabular}{c|ccc|c|c|c}
\hline
\multirow{2}{*}{Ref.} & \multicolumn{3}{c|}{Key components} & \multirow{2}{*}{\begin{tabular}[c]{@{}c@{}}Traffic\\ participants\end{tabular}} & \multirow{2}{*}{\begin{tabular}[c]{@{}c@{}}Transfer-\\ ability\end{tabular}} & \multirow{2}{*}{Objectives} \\
& Cloud & CAVs & RSUs & & & \\ \hline
{[9]} & $\surd$ & $\surd$ & × & CAV       & × & DT based AIoT RL system for lane changing \\ 
{[13]}  & × & $\surd$ & × & CAV          & × & Robust differential game for proceed at unsignalized intersections \\
{[14]}  & × & $\surd$ & × & CAV+HDV      & × & Game theoretic cooperative system for mixed traffic \\
{[15]}  & × & $\surd$ & × & CAV          & × & Virtual-spring distributed cooperative control method\\
{[18]}  & × & $\surd$ & × & CAV          & × & MARL with attention for unsignalized intersections \\
{[20]} & $\times$ & $\surd$ & $\times$ & CAV+HDV & $\times$ & Communication-driven MARL framework with intention sharing \\
{[22]}  & $\surd$ & $\surd$ & × & CAV+HDV   & × & Upstream platoon formation and scheduling system\\
{[25]} & $\surd$ & $\surd$ & $\surd$ & CAV+HDV & × & Infrastructure assisted imitation learning system \\
{[28]} & $\surd$ & $\surd$ & × & CAV       & $\surd$ & DT-assisted FRL system with fidelity guarantee \\
{[30]} & $\surd$ & $\surd$ & $\surd$ & CAV & $\times$ & DT-based predictive mobility analytics system for traffic management \\ \hline
\textbf{This work} & $\surd$ & $\surd$ & $\surd$ & CAV+HDV+VRUs & $\surd$ & Global \& local DT cooperative driving system with blind spot elimination\\ \hline
\end{tabular}
\begin{flushleft}
\footnotesize \textit{Note.} “Transferability” denotes validation of generalization to unseen intersection layouts.
\end{flushleft}
\end{table*}

Recent studies on cooperative driving at intersections can be broadly categorized into algorithm-oriented approaches and system-oriented approaches. Algorithm-oriented approaches emphasize modeling and optimizing the interactions among multiple agents. Game-theoretic approaches \cite{10440183,fang2023cooperativedrivingconnectedautonomous} aim to capture the interactive decision-making processes among CAVs. For instance, \cite{10440183} proposed a robust differential game model to address the dilemma of whether CAVs should proceed or yield at intersections. \cite{fang2023cooperativedrivingconnectedautonomous} combined k-level reasoning with cooperative games to improve coordination between CAVs and human-driven vehicles (HDVs). In addition, several physics-inspired algorithms have been proposed \cite{10265754,10976401,10704974} to more capture vehicle dynamics and interaction behaviors. Specifically, \cite{10265754} developed a distributed cooperative control algorithm based on a virtual spring system that achieves efficient and robust motion of CAVs. \cite{10976401} employed conflict-based search to jointly optimize CAV passing sequences and motion trajectories, aiming for energy-efficient operations at intersections, and \cite{10704974} integrated drivers’ visual characteristics into a risk-field model to more accurately model driving behavior and improve traffic efficiency at intersections. Meanwhile, deep reinforcement learning (DRL) and multi-agent reinforcement learning (MARL) have also been applied to cooperative driving \cite{11039015,10774177,11396022,10844516}. In particular, \cite{11039015,10774177} adopt self-attention mechanisms and combine discrete and continuous actions to learn coordination policies, thereby better adapting to complex and dynamic traffic environments. More recent studies \cite{11396022,10844516} have further considered autonomous driving in more realistic settings \cite{iqbal2019actorattentioncriticmultiagentreinforcementlearning}, including centralized training with decentralized execution \cite{11396022}, and communication-driven intention sharing \cite{10844516}, thus extending MARL to more realistic autonomous driving scenarios.

In contrast, system-oriented approaches emphasize the overall architecture and mechanisms for ITS. \cite{10251788} and \cite{10733743} are representative platoon-based studies. Specifically, \cite{10251788} proposed forming mixed platoons of CAVs and HDVs upstream and scheduling them at the platoon level to reduce coordination complexity, while \cite{10733743} analyzed platoon sizing strategies under diverse traffic demands to enhance intersection efficiency. Other approaches highlight the integration of infrastructure and signal control \cite{10571850} and \cite{YING2024104443}. Specifically, \cite{10571850} developed a cooperative lane assignment and longitudinal control strategy for intersections with contraflow lanes. Similarly, \cite{YING2024104443} proposed an infrastructure-assisted approach integrating imitation learning with coordinated signal timing and CAV trajectory control to improve efficiency. In addition, protocol-level solutions have been explored. For instance, \cite{10504747} proposed a distributed V2X-based intersection management system that improves robustness to violations and driving behaviors.

\begin{figure*}[t]
    \centerline{\includegraphics[width=0.9\textwidth]{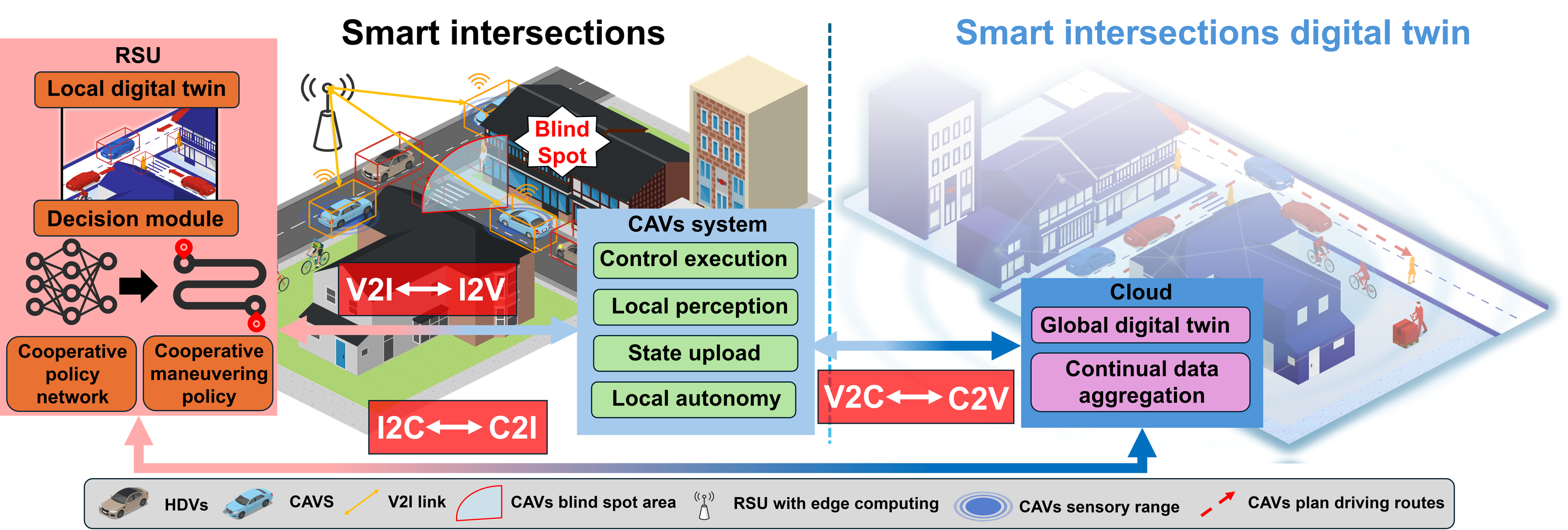}}
    \caption{High-level architecture of the DT-based V2I2V at smart intersections}
    \label{fig: CVIS}
\end{figure*}

More recently, DT technologies have been increasingly applied to cooperative driving studies, aiming to bridge the physical and cyber spaces while improving model fidelity and robustness\cite{DU20241666}. At the vehicle level, several studies \cite{10693297,11059924} have investigated DT-assisted multi-vehicle cooperative planning to enhance both safety and efficiency. For example, \cite{10693297} proposed a DT-assisted federated reinforcement learning (FRL) framework with predictive compensation for data loss to improve safety. Similarly, \cite{11059924} integrated artificial intelligence of things (AIoT) and DT technologies to develop a trajectory planning framework for lane-changing decisions and delays. At the system and infrastructure level, several studies \cite{10964767,10992265,11153748} have explored more advanced transportation applications. For example, \cite{10964767} focused on edge server deployment in ITS and proposed a two-stage resource allocation algorithm that combines spectral clustering with Q-learning to optimize latency, energy consumption, and load balancing. In addition, \cite{10992265} developed a DT framework for V2X-enabled vehicle corridors to support signal timing adjustment and vehicle advisory services. Furthermore, \cite{11153748} proposed a DT-based predictive mobility analytics system for real-time situational awareness and proactive traffic management.

Despite these advances, the integration of DT with cooperative driving at smart intersections remains limited. As summarized in Table \ref{tab:feature_compare}, most existing studies employ DT primarily for offline modeling and performance evaluation, rather than integrating it with cooperative decision-making algorithms. Although some studies utilize DT to enhance safety, they mainly focus on coordination among CAVs, instead of exploiting DT to construct a global perception that eliminates blind spots. In addition, prevailing decentralized vehicle-level decision-making paradigms often lead to locally optimal but system-inefficient behaviors, whereas our system enables globally optimized intersection-level performance. Moreover, vulnerable road users (VRUs) such as pedestrians and cyclists have received limited attention, even though they are the most exposed to blind-spot risks. In addition, the majority of studies rely on simulations or offline datasets for validation, with only a few exploring system implementations or real-world experiments to assess practical effectiveness. These limitations motivate the core contributions of this work.

\subsection{Contributions}

The main contribution of this paper is a novel DT-based vehicle-infrastructure-vehicle (V2I2V) cooperative system that enables bidirectional interaction between vehicles and RSUs for enhanced safety at smart intersections. Specifically, LiDAR-equipped RSUs construct a global BEV perception to generate a real-time digital replica of the intersection\cite{10584478}, providing situational perception beyond onboard sensing. Unlike most existing studies that emphasize only inter-vehicle cooperation, this paper incorporates pedestrians, cyclists, and HDVs into the cooperative model to improve safety for VRUs, who are most susceptible to blind spots\cite{sarlak2024enhancedcooperativeperceptionautonomous}. At the decision-making level, we propose a hierarchical reinforcement learning (HRL) framework. Offline pre-training is performed using conservative Q-learning (CQL) and behavior cloning (BC) to obtain a reliable initial policy from real-world datasets, followed by online fine-tuning with multi-agent proximal policy optimization (MAPPO) with self-attention mechanisms. This enables safe, efficient, and adaptive cooperative control among multiple CAVs. The integrated framework exhibits strong transferability and maintains excellent generalization performance even in unseen real-world intersection layouts\cite{bhatt2025architectingdigitaltwinsintelligent}. Finally, the system is validated through real-world proof-of-concept (PoC) trials. A deployment at the Institute of Science Tokyo campus confirms that the proposed system can eliminate blind spots, effectively protect VRUs, and significantly improve both traffic safety and efficiency. The main contributions of this paper can be summarized as follows: 

\begin{itemize}

\item \textbf{System architecture.} We design a novel V2I2V cooperative system that integrates RSUs and DT to eliminate blind spots and coordinated control at smart intersections.

\item \textbf{Multi-participant modeling.} We establish a unified cooperative system that explicitly incorporates pedestrians, cyclists, and HDVs into the decision-making loop.

\item \textbf{HRL-based control.} We propose a HRL framework that combines offline pre-training with online fine-tuning to achieve adaptive and robust cooperation among CAVs.

\item \textbf{Real-world validation.} We implement the proposed system in real-world PoC field trials on the Institute of Science Tokyo campus, demonstrating improved VRU safety and traffic performance compared with conventional autonomous driving systems (ADS).

\item \textbf{Scalability evaluation.} We evaluate the system across multiple real-world smart intersections reconstructed in simulation. Results show consistent performance gains over ADS baselines in both safety and efficiency under unseen layouts and diverse traffic conditions.

\end{itemize}

The rest of this paper is organized as follows. Section II introduces the architecture and workflow of the proposed V2I2V system. Section III presents the cooperative driving algorithms, including offline pre-training and online fine-tuning. Section IV describes the offline and online training processes, ablation studies, and multi-intersection simulations reconstructed from real Japanese layouts. Section V introduces the experimental platform, hardware and software configurations, and communication setup. Section VI reports the real-world PoC trials and quantitative evaluations. Section VII concludes the paper.

%% file: 2_SMDT.tex
This paper proposes a DT-based V2I2V cooperative system that integrates a cloud-level global DT with a RSU-level local DT, as shown in Fig.~\ref{fig: CVIS}. The purpose is to improve safety and efficiency at smart intersections, by decreasing risks from blind spots and enhancing protection for VRUs. The global DT maintains macroscopic traffic information, such as long-term flow statistics and risk maps, to support continuous system improvement, while local DT focuses on real-time environment modeling and cooperative control within RSU service area. Through this design, the system supports for collaboration among CAVs, HDVs, and VRUs.

To balance global consistency and local responsiveness, we design a global-local DT coordination mechanism. When vehicles are far from intersections, CAVs operate in onboard autonomous control modules, while the global DT provides data synchronization. As the CAV enters the RSU service area, the system automatically triggers the switch. In this area, blind spots and pedestrian flows reduce the reliability of onboard perception, while remote cloud-based control is constrained by latency and bandwidth fluctuations and cannot support real-time decision-making. By contrast, BEV and short-hop V2I communications enable the local DT to maintain comprehensive perception with low latency\cite{10978468}. The local DT then generates control signals by a decision-making model and transmits them to CAVs. After the vehicles pass through the intersection, the local DT summarizes abnormal events, such as emergency braking events, together with performance metrics such as safety margins, and stopping distance, and uploads them to the global DT to update statistical priors\cite{10049521}. Control subsequently returns to the vehicle after local DT control, forming a hierarchical closed-loop system that combines global consistency with local responsiveness\cite{10900364}.

In terms of communications architecture, the system adopts a dual-plane architecture\cite{9316767}. The control plane (C-plane) manages interactions between the global DT and CAVs before entering intersections. Meanwhile, the data plane (D-plane) supports high-bandwidth, low-latency exchanges between the local DT and CAVs within the RSU service area. The C-plane transmits macroscopic traffic guidance and status updates between the cloud and CAVs, whereas the D-plane ensures real-time exchange of sensing data and cooperative control signals between the RSU and CAVs. This layered communications design enables the architecture to maintain global consistency through the C-plane while providing low-latency response and cooperative CAV control through the D-plane. This dual-plane design can be incrementally integrated with existing ITS , which typically consists of traffic management centers, signal controllers, roadside sensing devices, and communication infrastructure. Specifically, the C-plane can support macroscopic traffic information exchange with traffic management centers, while the D-plane can interface with roadside sensing and RSU communication infrastructure to enhance intersection perception, blind spot elimination, and real-time cooperative decision-making without replacing conventional traffic monitoring and signal control functions.

At the decision-making level, the local DT deployed on the RSU adopts a HRL framework to generate cooperative control policies. Given the highly dynamic and partially observable nature of traffic environments, the problem is modeled as a partially observable Markov decision process (POMDP)\cite{ASTROM1965174}. RL is well suited for this problem, as it can optimize policies through interaction with the environment, handling uncertainty and multi-agent coordination. Specifically, the training process consists of two stages. Offline training with real-world traffic datasets produces reliable priors, while online fine-tuning in simulation environments enhances adaptability and robustness. Compared with rule-based or optimization-based ADS baselines, our approach naturally handles high-dimensional continuous action spaces while offering adaptability and generalization. Once deployed, the local DT executes the trained policies to provide centralized, real-time decision-making for CAVs, thereby significantly reducing onboard computational load and ensuring low-latency responses.

%% file: 4_implementation.tex
\subsection{Problem Formulation}

\begin{figure}[t]
    \centerline{\includegraphics[width=0.49\textwidth]{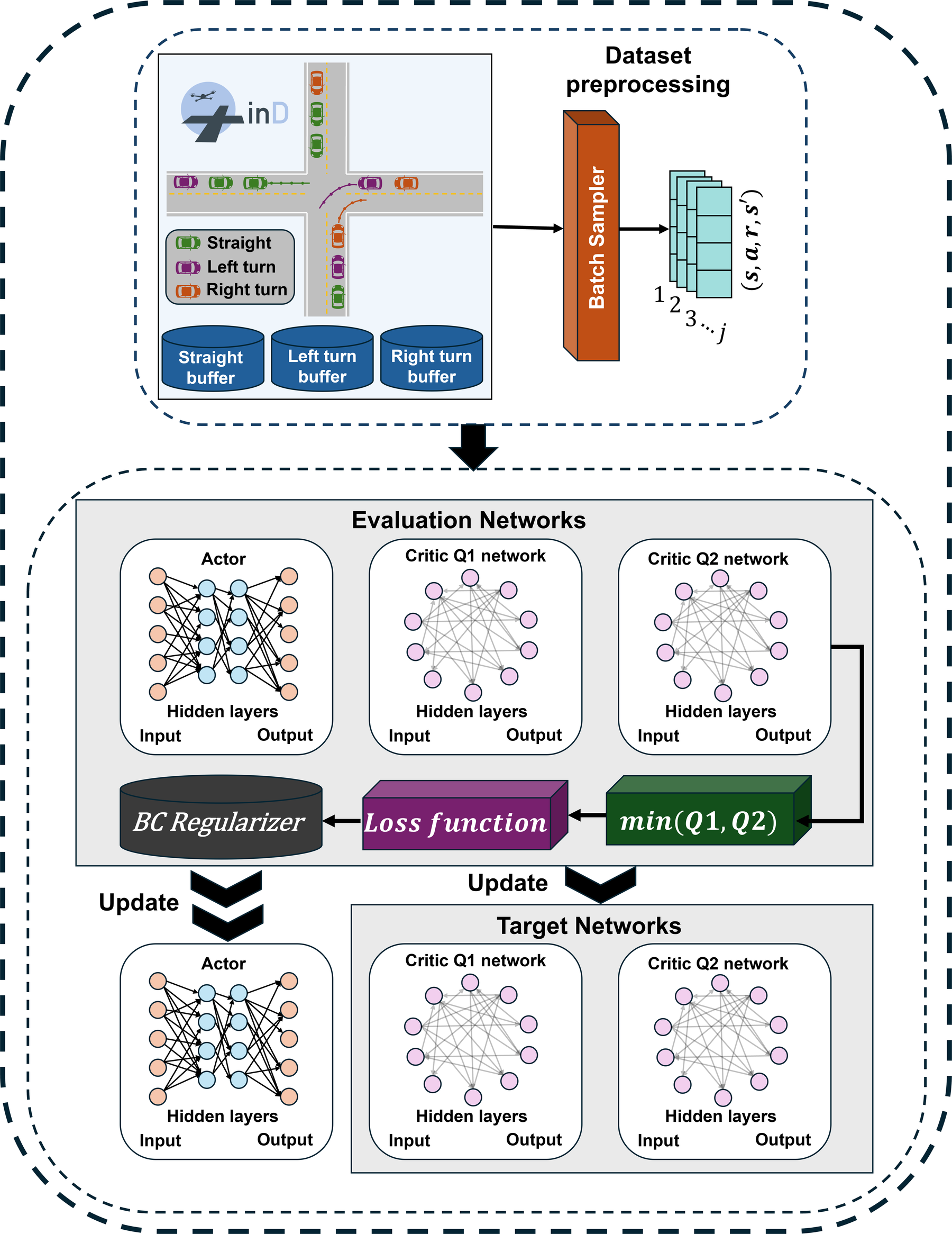}}
    \caption{Offline reinforcement learning pre-training with actor–critic networks}
    \label{fig: ofdl}
\end{figure}

We formulate a centralized cooperative decision-making problem for multiple CAVs at smart intersections. In this scenario, the RSU builds a comprehensive perception of intersection traffic by fusing its own sensing data with information shared by CAVs via V2X communications. Under this centralized control system, the RSU generates control signals for all CAVs within its service area, enabling cooperative driving.

We model the cooperative decision-making problem as a POMDP:
\begin{equation}
\mathcal{G}=\langle \mathcal{N}, \mathcal{S}, \mathcal{O}_c, \mathcal{A}, \mathcal{P}, r, \gamma \rangle .
\end{equation}
Here, $\mathcal{N}=\{1,\dots,N\}$ represents the set of CAVs, $\mathcal{S}$ is the global traffic state, and $\mathcal{O}_c$ is the centralized observation constructed by the local DT at the RSU. The joint action space is $\mathcal{A}=\mathcal{A}_1\times\cdots\times\mathcal{A}_N$. At each time step $t$, the joint action is $a^t=(a_1^t,\dots,a_N^t)$ with $a_i^t\in\mathcal{A}_i$. The transition kernel is $\mathcal{P}(s^{t+1}\mid s^t,a^t)$,which specifies the distribution of the next state $s^{t+1}$ given the current state $s^t$ and joint action $a^t$. The reward function is $r:\mathcal{S}\times\mathcal{A}\to\mathbb{R}$, quantifying safety and efficiency for the joint action, and $\gamma\in[0,1]$ is the discount factor.

The RSU executes a centralized policy $\pi_c(a^t \mid o_c^t)$, which produces a joint action distribution over $\mathcal{A}$ conditioned on the centralized observation $o_c^t\in\mathcal{O}_c$. The objective is to maximize the expected discounted return:

\begin{equation}
\max_{\pi_c}~ J(\pi_c)=\mathbb{E}_{\tau\sim \pi_c}\!\left[\sum_{t=0}^{T}\gamma^t r^t\right],
\end{equation}

where $T$ is the decision horizon, $\tau$ is a trajectory generated under $\pi_c$, and $r^t=r(s^t,a^t)$ is the instantaneous reward at time $t$. This formulation enables the RSU to learn a cooperative policy that balances short-term safety with long-term traffic efficiency.

\begin{figure*}[!t]
    \centerline{\includegraphics[width=0.9\textwidth]{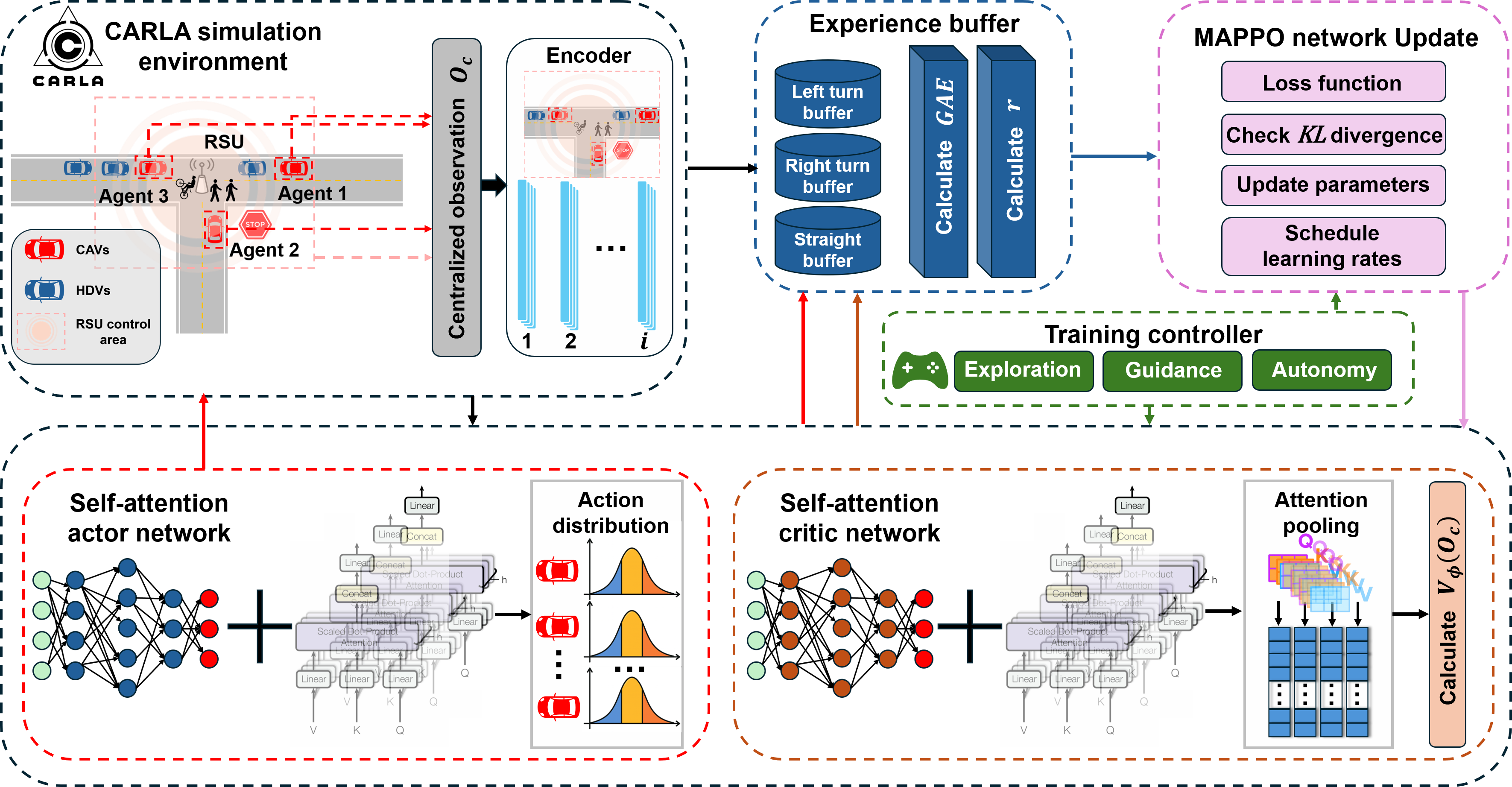}}
    \caption{Online reinforcement learning fine-tuning with actor–critic networks}
    \label{fig: onRL}
\end{figure*}

\subsection{Observation and Action Modeling}

At each time step $t$, the RSU aggregates information uploaded by vehicles and fuses it with its own sensor data to construct a global perception of the intersection, formulated as
\begin{equation}
o_c^t = \mathcal{F}_o\big(y_{\text{RSU}}^t,\{y_i^t\}_{i\in\mathcal{N}}\big) \in \mathcal{O}_c,
\end{equation}
where $y_{\text{RSU}}^t$ denotes the RSU’s sensor data and $y_i^t$ is the perception information transmitted by vehicle $i$ via V2X.

The local observation of each vehicle $i$ is represented as a fixed-length feature vector $o_i^t$, composed of three semantically distinct groups:
\begin{equation}
o_i^t = \big[\, o^{t}_{i,\text{self}},\; o^{t}_{i,\text{rel}},\; o^{t}_{i,\text{road}} \,\big],
\end{equation}
where the feature groups are defined as follows:

\begin{itemize}
    \item \textbf{Ego-vehicle kinematics} $o^{t}_{i,\text{self}}$: longitudinal velocity, position in the local coordinate frame, heading angle, distance to the conflict area, a binary indicator of intersection occupancy, and the relative position to the most critical neighboring vehicle.
    
    \item \textbf{Inter-vehicle interactions} $o^{t}_{i,\text{rel}}$: relative longitudinal velocity to the nearest vehicle, minimum inter-vehicle distance, estimated time-to-collision (TTC), speed limit of the current road segment, and the number of surrounding agents.
    
    \item \textbf{Road geometry and navigation} $o^{t}_{i,\text{road}}$: heading error and lateral offset relative to the next waypoint, distance to the destination, road curvature, right-of-way indicator, and stopping distance margin.
\end{itemize}

The low-level control action of each vehicle is a two-dimensional continuous vector:
\begin{equation}
a_i^t = \big[a^{t}_{i,\text{long}},\; a^{t}_{i,\text{lat}}\big],
\end{equation}
where $a^{t}_{i,\text{long}}$ corresponds to longitudinal control and $a^{t}_{i,\text{lat}}$ corresponds to lateral control.

\subsection{Reward Function}

We design a composite reward function that combines multiple sub-rewards to guide the learning of safe and efficient intersection management policies. Each sub-reward component is weighted by an independent scaling factor $\alpha_k$ to balance its contribution during training. They were determined through preliminary empirical tuning across repeated training runs, with priority given to safety-related objectives, especially collision avoidance, followed by task completion and traffic efficiency. The overall reward function is formulated as:
\begin{equation}
r = r_{\text{success}} + r_{\text{progress}} + r_{\text{safety}} + r_{\text{collision}} + r_{\text{smooth}}.
\end{equation}

\begin{itemize}

\item \textbf{Task completion reward:}  
A positive reward is granted when all vehicles successfully traverse the intersection:
\begin{equation}
r_{\text{success}} =
\begin{cases}
+\alpha_{\text{succ}}, & \text{all CAVs reach their destinations}, \\
0, & \text{otherwise}.
\end{cases}
\end{equation}

\item \textbf{Progress reward:}  
To encourage continuous advancement, each vehicle is rewarded according to its progress along the planned trajectory:
\begin{equation}
r_{\text{progress}} = \alpha_p \cdot \sum_{i=1}^N \Delta d_i,
\end{equation}
where $\Delta d_i$ is the displacement of vehicle $i$ projected onto its route during the current step.

\item \textbf{Safety reward:}  
Risky interactions are penalized using the time-to-collision (TTC):
\begin{equation}
\text{TTC}_{ij} = \frac{d_{ij}}{\max(v_{rel}^{ij}, \epsilon)},
\end{equation}
where $d_{ij}$ is the relative distance between vehicles $i$ and $j$, $v_{rel}^{ij} = |v_i - v_j|$ their relative velocity, and $\epsilon$ is a small constant for numerical stability.  
The penalty is defined as
\begin{equation}
r_{\text{safety}} = -\alpha_s \cdot \sum_{(i,j) \in \mathcal{P}} \exp\!\left(-\frac{\text{TTC}_{ij}}{\beta}\right),
\end{equation}
where $\mathcal{P}$ is the set of interacting vehicle pairs within a predefined safety radius, and $\beta$ controls the penalty sensitivity. This formulation imposes larger penalties to smaller TTC while diminishing effects for safe distances.

\item \textbf{Collision Penalty:}  
Severe penalties are imposed for collision events:
\begin{equation}
r_{\text{collision}} =
\begin{cases}
-\alpha_c, & \text{if a collision occurs}, \\
0, & \text{otherwise}.
\end{cases}
\end{equation}

\item \textbf{Smoothness reward:}  
To encourage smooth and comfortable driving behaviors, abrupt changes in actions are penalized. Let $\Delta a_i^t = a_i^t - a_i^{t-1}$ represent the temporal variation in the control command of vehicle $i$, then the smoothness reward is defined as:
\begin{equation}
r_{\text{smooth}} = -\alpha_{\text{sm}} \cdot \sum_{i=1}^N \|\Delta a_i^t\|^2,
\end{equation}
where $\|\Delta a_i^t\|^2$ measures the magnitude of action variation between consecutive steps.  
\end{itemize}

\subsection{Offline Pre-training: Networks and Algorithm}

\begin{algorithm}[t]
\caption{Offline Pre-training}
\label{alg:offline}
\SetAlgoLined
\KwIn{$\mathcal{D}_{\text{InD}}$ // Offline dataset\\
      $M_{\text{offline}}$ // Number of offline training epochs}
\KwOut{$\theta^*$ // Pre-trained policy parameters}

Initialize $\pi_\theta(a|x)$, $Q_{\phi_1}(x,a)$, $Q_{\phi_2}(x,a)$, $\bar{Q}_{\phi_1}$, $\bar{Q}_{\phi_2}$\;; Classify $\mathcal{D}_{\text{InD}}$ into buffers: $\mathcal{D}_{\text{left}}$, $\mathcal{D}_{\text{right}}$, $\mathcal{D}_{\text{straight}}$\;.

\For{$epoch = 1$ to $M_{\text{offline}}$}{
    $\mathcal{B} \leftarrow$ sample\_batch($\mathcal{D}_{\text{InD}}$, batch\_size)\;;
    
    \For{$(x, a, r, x', done) \in \mathcal{B}$}{
        // Compute TD target using Eq. (12)\;.\\
        $y \leftarrow r + \gamma(1-done) \cdot \min(\bar{Q}_{\phi_1}(x',a'), \bar{Q}_{\phi_2}(x',a'))$, where $a' \sim \pi_\theta(x')$\;;
        
        \For{$i \in \{1,2\}$}{
            Compute $\mathcal{L}_{\text{CQL}}$ using Eq. (13) with $\alpha$ and $a_{\text{data}}$\;,
            $\phi_i \leftarrow \phi_i - \eta\nabla_{\phi_i} \mathcal{L}_{\text{CQL}}$\;;
        }
        
        // Update policy with BC regularization (Eq. 14)\;.\\
        Compute $\mathcal{L}_\pi$ using Eq. (14) with $\lambda$\;,
        $\theta \leftarrow \theta - \eta\nabla_\theta \mathcal{L}_\pi$\;;
        
        // Soft update target networks\;.\\
        $\bar{Q}_{\phi_i} \leftarrow \tau Q_{\phi_i} + (1-\tau)\bar{Q}_{\phi_i}$\;;
    }
}
\end{algorithm}

During the offline training phase, we construct a replay buffer $\mathcal{D}$ based on the InD real-world dataset\cite{bock2019inddatasetdronedataset}, and partition the driving behaviors at smart intersections into three sub-buffers: left turn, straight, and right turn. This design allows the model to learn decision-making features under distinct maneuver intentions, enhancing its generalization across intersection scenarios. To learn stable and transferable policies from offline dataset, we adopt an actor–critic framework. Furthermore, we combine the joint optimization of CQL and BC to simultaneously improve policy stability and ensure closeness to the expert distribution\cite{kumar2020conservativeqlearningofflinereinforcement,NIPS1988_812b4ba2}, as shown in Fig.~\ref{fig: ofdl}. The detailed training procedure is summarized in Algorithm~\ref{alg:offline}. With this design, the model can learn stable and transferable policies. Although the InD dataset may introduce some dataset bias, it is used mainly to learn basic driving knowledge, while the policy is subsequently adapted and refined through online learning under Japanese traffic rules. Ultimately, the learned policy can not only serve as a warm start for online training, but also naturally extend to an arbitrary number of CAVs, facilitating subsequent fine-tuning across different environments. The detailed training hyperparameters are listed in Table \ref{tab:key_hyperparameters}.

\begin{table}[t]
\centering
\caption{Key hyper-parameters used in training}
\label{tab:key_hyperparameters}
\begin{tabular}{c|c}
\hline
\textbf{Category} & \textbf{Parameter / Value} \\ \hline
\begin{tabular}[c]{@{}c@{}}Reward\\ coefficients\end{tabular} & 
\begin{tabular}[c]{@{}c@{}}$\alpha_{\mathrm{succ}}=100.0$,\quad $\alpha_p=0.5$,\quad $\alpha_s=2.0$\\ $\beta=3.0$,\quad $\alpha_c=50.0$,\quad $\alpha_{sm}=0.1$\end{tabular} \\ \hline
\begin{tabular}[c]{@{}c@{}}Training\\ parameters\end{tabular} & 
\begin{tabular}[c]{@{}c@{}}$\gamma=0.99$,\quad $\lambda=0.95$,\quad $\epsilon=0.1$\end{tabular} \\ \hline
\begin{tabular}[c]{@{}c@{}}Self-attention\\ parameters\end{tabular} & 
\begin{tabular}[c]{@{}c@{}}Heads: 8,\quad Layers: 3,\quad Hidden dim: 256\\ Dropout: 0.1,\quad Position encoding: learned\end{tabular} \\ \hline
\end{tabular}
\end{table}

For the critic network, the update is based on a combination of the temporal-difference (TD) target and the CQL regularizer. Given a state–action pair, the target is defined as:

\begin{equation}
y = r + \gamma \min(\bar{Q}_{{\theta}_1}(x',a'), \bar{Q}_{{\theta}_2}(x',a')), \quad a' \sim \pi_\theta(x'),
\end{equation}

where $r$ is the immediate reward, $\gamma \in (0,1)$ is the discount factor, $x'$ is the next state, and $a'$ is the action generated by the current policy $\pi_\theta$, $\bar{Q}_{{\theta}_1}$ and $\bar{Q}_{{\theta}_2}$ are two target Q networks. The critic loss is then:

\begin{equation}
\begin{aligned}
\mathcal{L}_{\text{CQL}} 
= & \; \mathbb{E}_{(x,a,r,x')\sim \mathcal{D}} 
    \left[ (Q_{\theta_i}(x,a) - y)^2 \right] \\
& + \alpha \, \mathbb{E}_{x \sim \mathcal{D}} 
    \left[ \log \sum_{a_j} e^{Q_{\theta_i}(x,a_j)} 
    - Q_{\theta_i}(x,a_{\text{data}}) \right],
\end{aligned}
\end{equation}

Here, $Q_\theta(x,a)$ represents the current Q-network, $y$ is the TD target, $\alpha > 0$ is the weight of the regularizer, and $\pi_\beta$ represents the expert behavior policy. The first term is the TD error and is used to learn the value function. The second term is the CQL regularizer, where $a_j$ includes random actions sampled from a uniform distribution as well as actions generated by the current policy, and $a_{\text{data}}$ is the expert action from the dataset. This regularizer penalizes the Q-values of out-of-distribution actions, thereby ensuring that the policy remains conservative in regions with insufficient data support.

For the actor network, the optimization objective considers both maximizing Q-values and maintaining consistency with the expert distribution:

\begin{equation}
\begin{aligned}
\mathcal{L}_{\pi} 
= & - \mathbb{E}_{x\sim \mathcal{D}} \Big[ 
      \min \big( Q_{\theta_1}(x, \pi_\theta(x)), 
                 Q_{\theta_2}(x, \pi_\theta(x)) \big) \Big] \\
& + \lambda \, \mathbb{E}_{(x,a)\sim \mathcal{D}} 
    \big[ \|\pi_\theta(x)-a\|^2 \big],
\end{aligned}
\end{equation}

where the first term increases the policy value, and the second term is the BC regularizer, constraining the learned policy $\pi_\theta$ to stay close to the expert actions $a$, $\lambda > 0$ controls the regularization strength.

The final joint optimization objective is:
\begin{equation}
\mathcal{L} = \mathcal{L}_{\text{CQL}}^{Q_1} + \mathcal{L}_{\text{CQL}}^{Q_2} + \mathcal{L}_{\pi}
\end{equation}

\subsection{Online Fine-tuning: Networks and Algorithm}

After completing the offline pre-training phase, we further conduct online training in the CARLA simulation environment\cite{dosovitskiy2017carlaopenurbandriving}. A digital replica of the target PoC intersection is built in CARLA and used as the online RL training environment. In this process, the policy starts from offline initialized parameters and no longer relies on fixed expert demonstrations. Instead, it continuously updates through interactions with the dynamic environment, thereby enhancing its adaptability to unseen intersection scenarios. The overall online fine-tuning framework is illustrated in Fig.~\ref{fig: onRL}.  

We adopt the MAPPO algorithm as the core framework\cite{yu2022surprisingeffectivenessppocooperative}. The overall online fine-tuning process based on MAPPO is summarized in Algorithm~\ref{alg:online}. At each time step $t$, the RSU generates an action $a^t$ based on centralized observation $o_c^t$, while the critic network estimates its value $V_\phi(o_c^t)$. To improve sample efficiency while balancing bias and variance, we employ generalized advantage estimation (GAE)\cite{schulman2018highdimensionalcontinuouscontrolusing}, defined as:  
\begin{equation}
\hat A_t = \sum_{l=0}^{T-1-t} (\gamma\lambda)^l \Big(r_{t+l} + \gamma V_\phi(o_c^{\,t+l+1}) - V_\phi(o_c^{\,t+l}) \Big),
\end{equation}
where $\gamma \in (0,1]$ is the discount factor and $\lambda \in [0,1]$ is the GAE coefficient.

Based on the advantage function, the policy optimization objective follows the PPO clipped loss:

\begin{equation}
\begin{aligned}
L^{\text{PPO}}(\theta)
= {} & \mathbb{E}_t \Big[ \min \big(
r_t(\theta)\,\hat A_t, \\
& \mathrm{clip}\big(r_t(\theta),\,1-\varepsilon,\,1+\varepsilon\big)\,\hat A_t
\big) \Big].
\end{aligned}
\end{equation}

where  
\begin{equation}
r_t(\theta) = \frac{\pi_\theta(a^t|o_c^t)}{\pi_{\theta_{\text{old}}}(a^t|o_c^t)}
\end{equation}
is the policy ratio and $\varepsilon$ is the clipping threshold.

The critic is updated by minimizing the mean squared error:  
\begin{equation}
L^{\text{critic}}(\phi) = \mathbb{E}_t\!\left[\big(V_\phi(o_c^t)-\hat V_t\big)^2\right],
\end{equation}
where $\hat V_t$ is the target value derived from advantage backtracking.  

Specifically, the critic network processes the centralized observation through stacked self-attention layers, followed by attention pooling to output a scalar value:  
\begin{equation}
V_\phi(o_c^t) = \text{Linear}(\text{AttentionPool}(\text{SA}^3(\text{Encoder}(o_c^t)))),
\end{equation}
where $\text{SA}^3$ represents three stacked self-attention blocks, and AttentionPool aggregates multi-vehicle features into a global state representation.  

To better capture multi-vehicle interactions, we introduce self-attention mechanisms into both the actor and critic networks. Let $h_i^{(l)}$ denotes the hidden state of vehicle $i$ at layer $l$, with $H^{(l)}$ representing the set of all hidden states. The update rule is given by:  
\begin{equation}
h_i^{(l+1)} = \text{LN}\big(h_i^{(l)} + \text{MHAttn}(h_i^{(l)}, H^{(l)}, H^{(l)})\big),
\end{equation}
where $\text{MHAttn}$ represents multi-head attention and LN represents layer normalization. Although the RL-based controller is not presented as a fully interpretable model, the self-attention mechanism partially improves transparency by modeling inter-agent relevance in the decision process. The resulting attention weights provide a useful cue for analyzing which surrounding vehicles or traffic interactions are emphasized by the controller in different situations.

\begin{algorithm}[t]
\caption{Online Fine-tuning}
\label{alg:online}
\SetAlgoLined
\KwIn{$\theta^*$ // Pre-trained policy parameters \\
      Env // Online environment \\
      $M_{\text{online}}$ // Number of online episodes}
\KwOut{$\psi^*$, $\phi^*$ // Optimized policy and value networks}

Initialize Actor $\pi_\psi$ and Critic $V_\phi$\;; Load pre-trained encoder weights from $\theta^*$ into $\pi_\psi$\;; Initialize buffer $\mathcal{B} \leftarrow \emptyset$ and define curriculum (Eq. 24)\;.

\For{$episode = 1$ to $M_{\text{online}}$}{
    Load phase parameters $\eta_t$ from Eq. (24)\;,\\
    $N_t \sim$ Uniform$(1, N_{\max})$, $\mathcal{O} \leftarrow$ Env.reset$(N_t)$\;;
    
    \For{$t = 1$ to $T_{\max}$}{
    // Action selection following curriculum policy (Eq. 24). \\
        Sample $a_i$ from $\pi_{\text{curriculum}}(\mathcal{O}, \eta_t)$ for $i=1,\ldots,N_t$\;;\\
        $\mathcal{O}', \mathcal{R}, done \leftarrow$ Env.step$(\mathcal{A})$\;;\\
        $\mathcal{B}$.add$(\mathcal{O}, \mathcal{A}, \mathcal{R}, V_\phi(\mathcal{O}), \log\pi, \mathcal{M})$\;
        $\mathcal{O} \leftarrow \mathcal{O}'$\;;\\
        \If{done}{break\;;}
    }
    
    // MAPPO Update. \\
    \If{$|\mathcal{B}| \geq$ batch\_size}{
        Compute GAE $\hat{A}_t$ using Eq. (17), normalize $\hat{A}$\;;
        
        \For{$k = 1$ to $K_{\text{epochs}}$}{
            \For{mini\_batch $\in \mathcal{B}$.sample()}{
                Compute $\mathcal{L}^{\text{PPO}}$, $\mathcal{L}^{\text{critic}}$, $\mathcal{L}_{\text{entropy}}$ (Eq. 18-20, 24)\;,\\
                $\mathcal{L} \leftarrow \mathcal{L}^{\text{PPO}} + c_{\text{value}} \mathcal{L}^{\text{critic}} + \mathcal{L}_{\text{entropy}}$\;;\\
                Update $\psi \leftarrow \psi - \eta_{\text{actor}} \nabla_\psi \mathcal{L}$,\\ 
                $\phi \leftarrow \phi - \eta_{\text{critic}} \nabla_\phi \mathcal{L}^{\text{critic}}$\;;\\
                \If{KL$(\pi_{\text{old}} \| \pi_{\text{new}}) > \delta_t$}{break\;}
            }
        }
        $\mathcal{B}$.clear()\;
    }
    
    Update $\delta_t$ (Eq. 25), $\lambda_t$ (Eq. 26)\;.
}

\end{algorithm}

To enhance policy generalization, the number of participating vehicles during training is dynamically varied as $N_t \sim \text{Uniform}(1, N_{\max})$. The attention mechanism is adapted to handle variable agent sets by applying a binary mask $\mathcal{M}$ over inactive agents in the attention computation:
\begin{equation} 
\text{Attention}(\mathbf{Q},\mathbf{K},\mathbf{V}) = 
\text{softmax}\!\left(\frac{\mathbf{Q}\mathbf{K}^\top}{\sqrt{d_k}} + \mathcal{M}_{\text{mask}}\right)\mathbf{V},
\end{equation}
where $\mathbf{Q}$, $\mathbf{K}$, and $\mathbf{V}$ denote the query, key, and value matrices, respectively, $d_k$ is the key dimension, and $\mathcal{M}_{\text{mask},ij}=-\infty$ for masked agents.

To further stabilize training, we introduce a three-stage curriculum learning scheme. The training is divided into exploration, guided, and autonomous phases, with hyperparameters adjusted accordingly:   
\begin{equation}
\eta_t = 
\begin{cases}
\eta_1, & t < T_1,\\
\eta_2, & T_1 \le t < T_2,\\
\eta_3, & t \ge T_2.
\end{cases}
\end{equation}
where $\eta_1$, $\eta_2$, and $\eta_3$ correspond to the exploration phase, the guided learning phase with moderate exploration, and the autonomous phase focusing on stable policy convergence.

Correspondingly, the KL divergence threshold is dynamically adjusted:
\begin{equation}
\delta_t=
\begin{cases}
5\delta_0, & \text{exploration},\\
2\delta_0, & \text{guided},\\
\delta_0 \cdot f(\rho), & \text{autonomous},
\end{cases}
\end{equation}
where $\delta_0$ is the base threshold and $\rho$ denotes the success rate, with $f(\rho)$ mapping it to the adjustment factor.

Furthermore, the learning rate is scaled adaptively based on the success rate:  
\begin{equation}
\lambda_t = \lambda_0 \cdot (1 + \kappa \cdot SR_t),
\end{equation}
where $SR_t$ is the success rate within a sliding window, $\lambda_0$ is the initial learning rate, and $\kappa$ is an adaptive coefficient.

\begin{figure*}[!t]
    \centering
    \begin{minipage}[b]{0.32\textwidth}
        \centering
        \includegraphics[width=\textwidth]{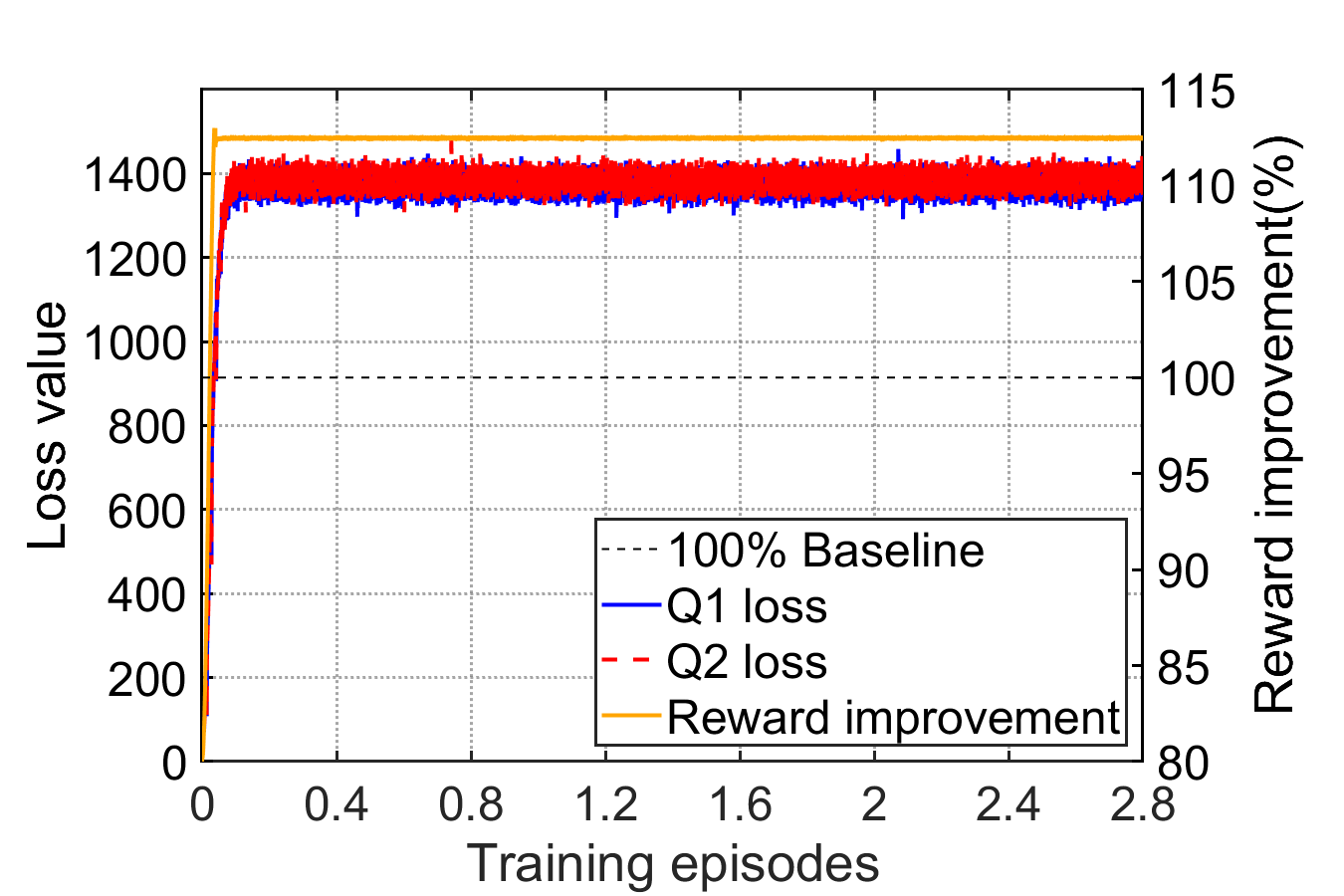}
        \centerline{(a)}
    \end{minipage}
    \hfill
    \begin{minipage}[b]{0.32\textwidth}
        \centering
        \includegraphics[width=\textwidth]{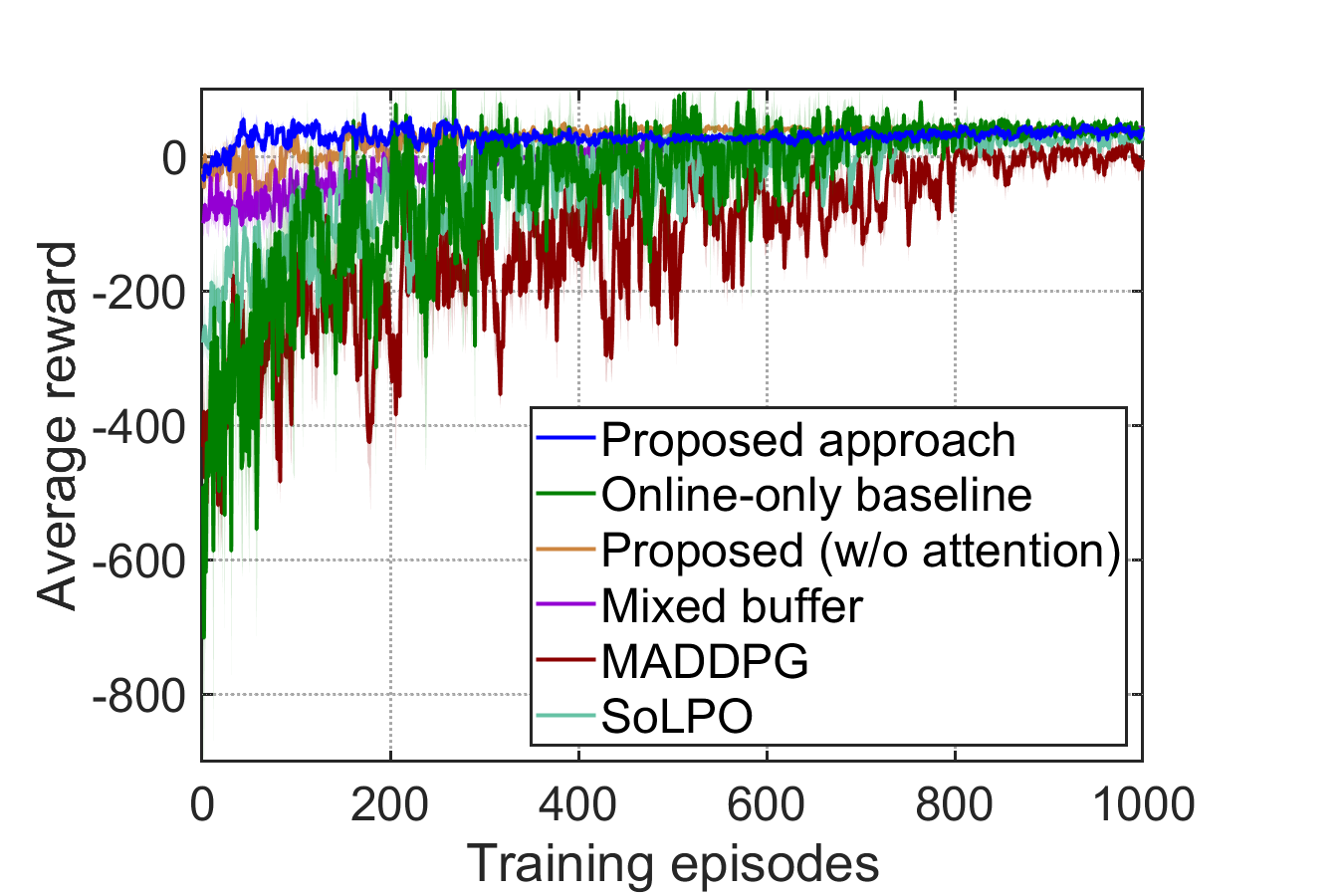}
        \centerline{(b)}
    \end{minipage}
    \hfill
    \begin{minipage}[b]{0.32\textwidth}
        \centering
        \includegraphics[width=\textwidth]{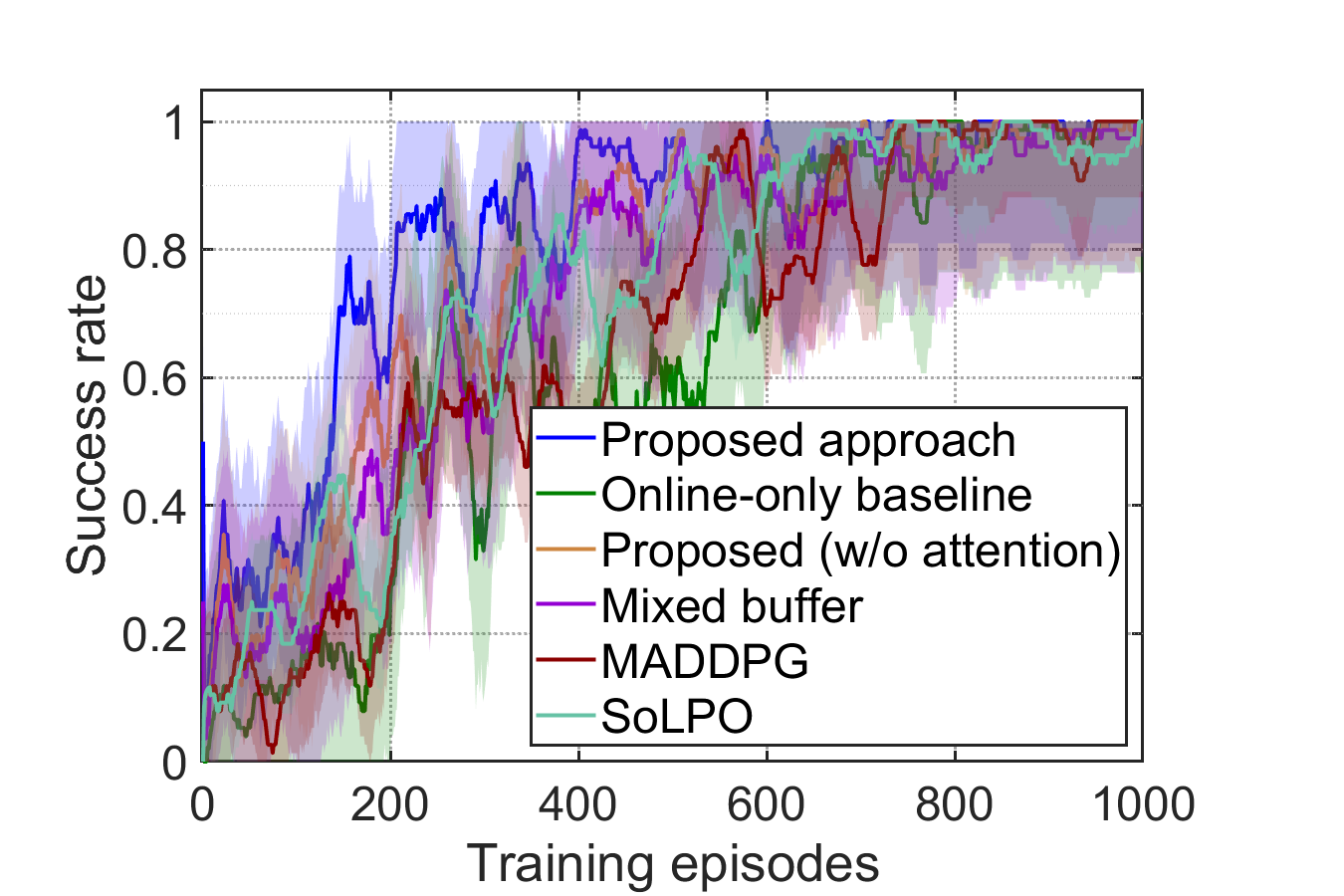}
        \centerline{(c)}
    \end{minipage}
    \caption{(a) Offline pre-training results; (b) comparison of training performance across different approaches in terms of average reward; (c) success rate over training episodes}
    \label{res}
\end{figure*}

Overall, this online training approach builds on MAPPO by integrating self-attention for multi-agent interaction modeling, reward shaping with domain priors, staged curriculum learning, and dynamic KL regularization. With this design, the policy learns stable and efficient control strategies in complex traffic scenarios, narrowing the gap between simulation and real-world deployment.

%% file: 5_Sim.tex
\subsection{RL Training and Ablation Study}


We conduct comprehensive offline and online experiments to verify the effectiveness of the proposed HRL framework. In the offline pre-training stage, we present the training results on the InD dataset, including the Q1/Q2 loss curves and normalized rewards, to demonstrate the training stability and the performance gains brought by the combination of CQL and BC. In the online training stage, we fine-tune the policy in the CARLA simulator and perform both ablation studies and comparisons with representative SOTA methods. The evaluation compares the proposed method in terms of average reward, success rate, and convergence speed.

\begin{figure}[t]
    \centerline{\includegraphics[width=0.49\textwidth]{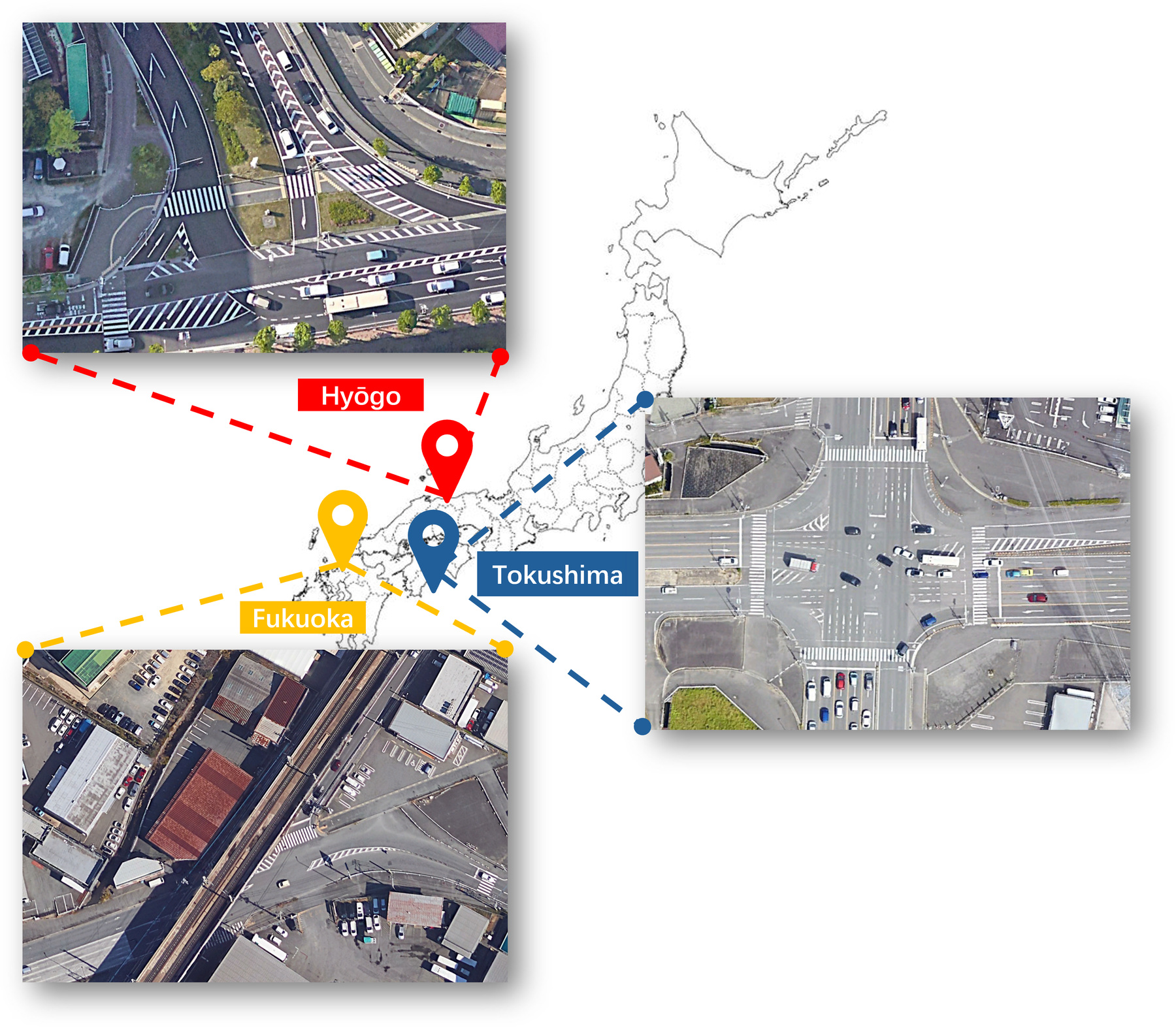}}
    \caption{Locations and aerial views of the three representative smart intersections}
    \label{m}
\end{figure}

Fig.~\ref{res} (a) presents the Q1/Q2 loss curves and normalized reward improvement during offline training. Both Q-function losses converge smoothly, indicating stable value learning. The normalized reward reaches approximately 112\% of the expert level, showing that CQL combined with BC not only reproduces expert behavior but also exceeds expert performance. This enhanced initialization significantly reduces the exploration required during online fine-tuning, leading to faster convergence and improved training stability.

\begin{figure*}[t]
    \centering
    \begin{minipage}[b]{0.32\textwidth}
        \centering
        \includegraphics[width=\textwidth]{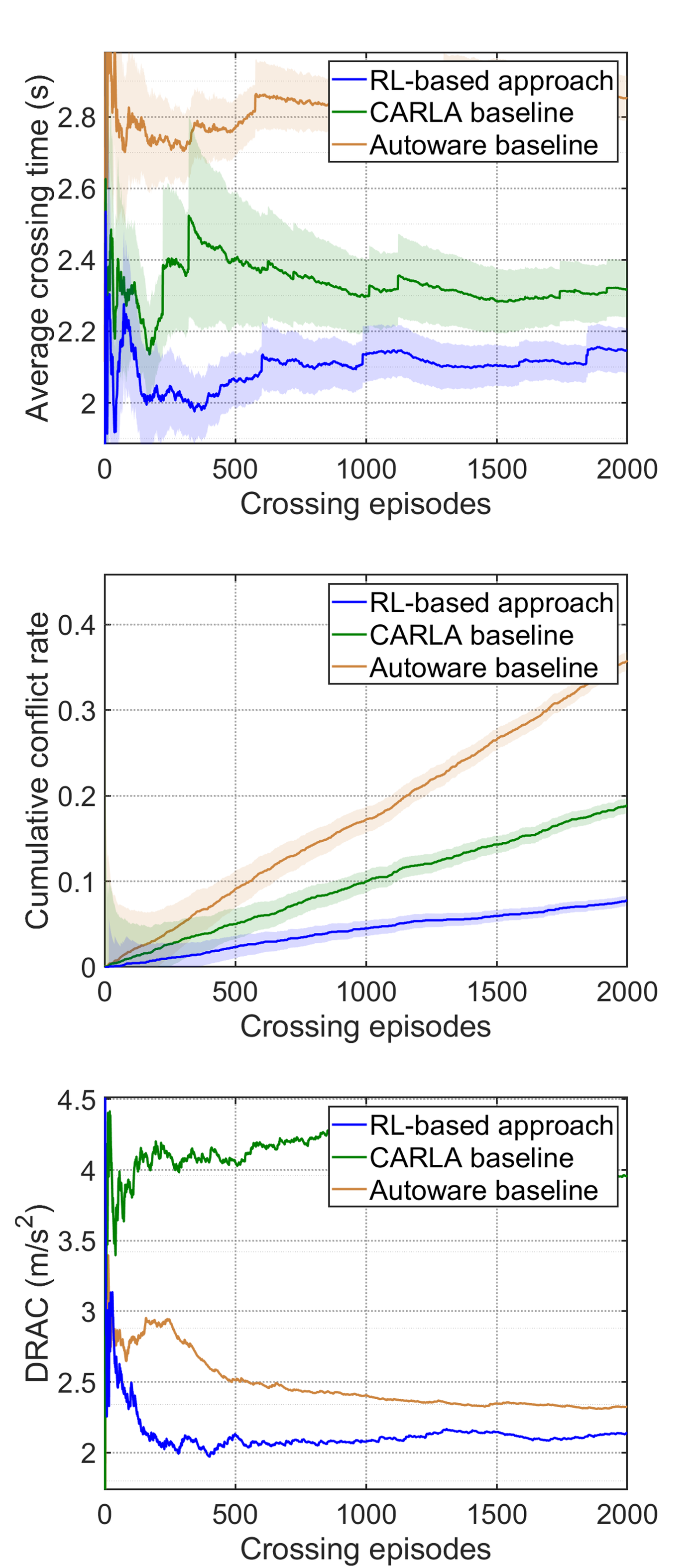}
        \centerline{(a) Hyogo}
    \end{minipage}
    \hfill
    \begin{minipage}[b]{0.32\textwidth}
        \centering
        \includegraphics[width=\textwidth]{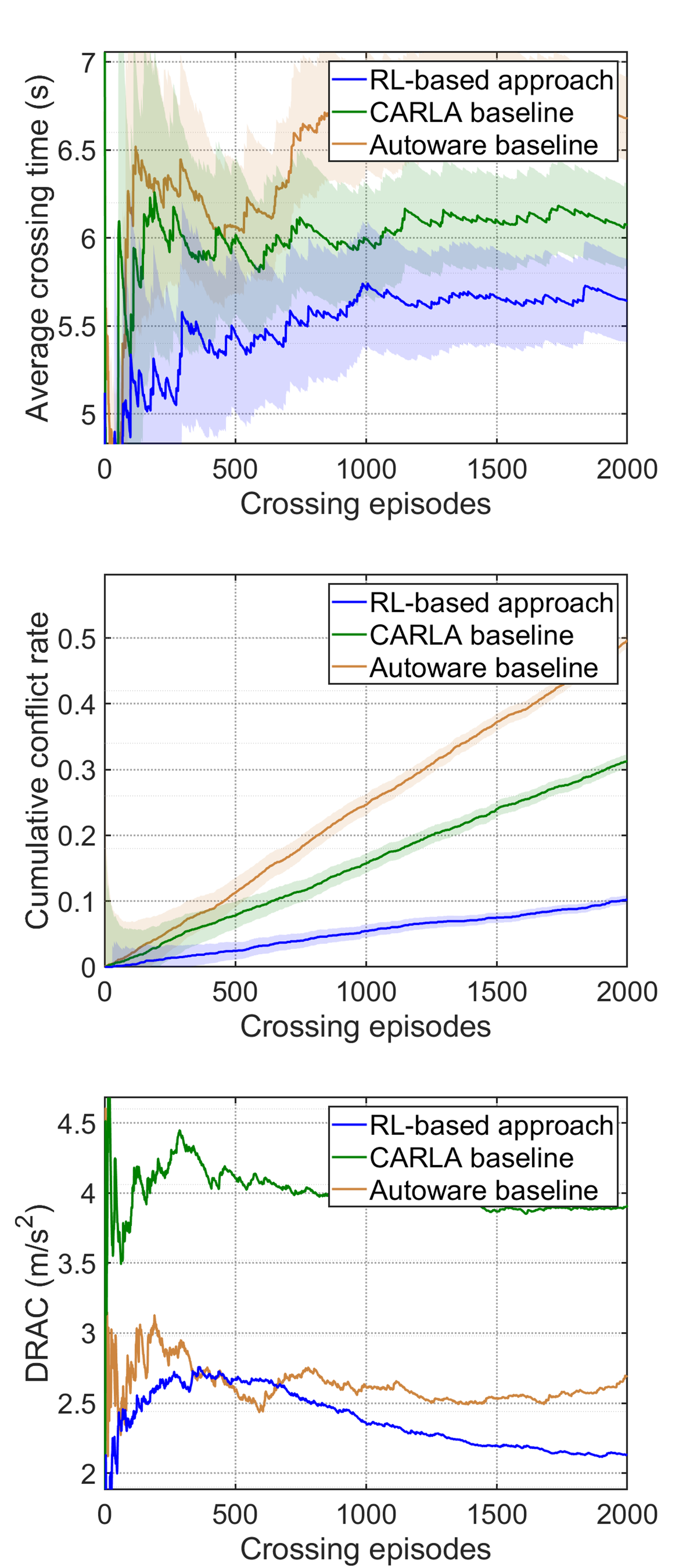}
        \centerline{(b) Fukuoka}
    \end{minipage}
    \hfill
    \begin{minipage}[b]{0.32\textwidth}
        \centering
        \includegraphics[width=\textwidth]{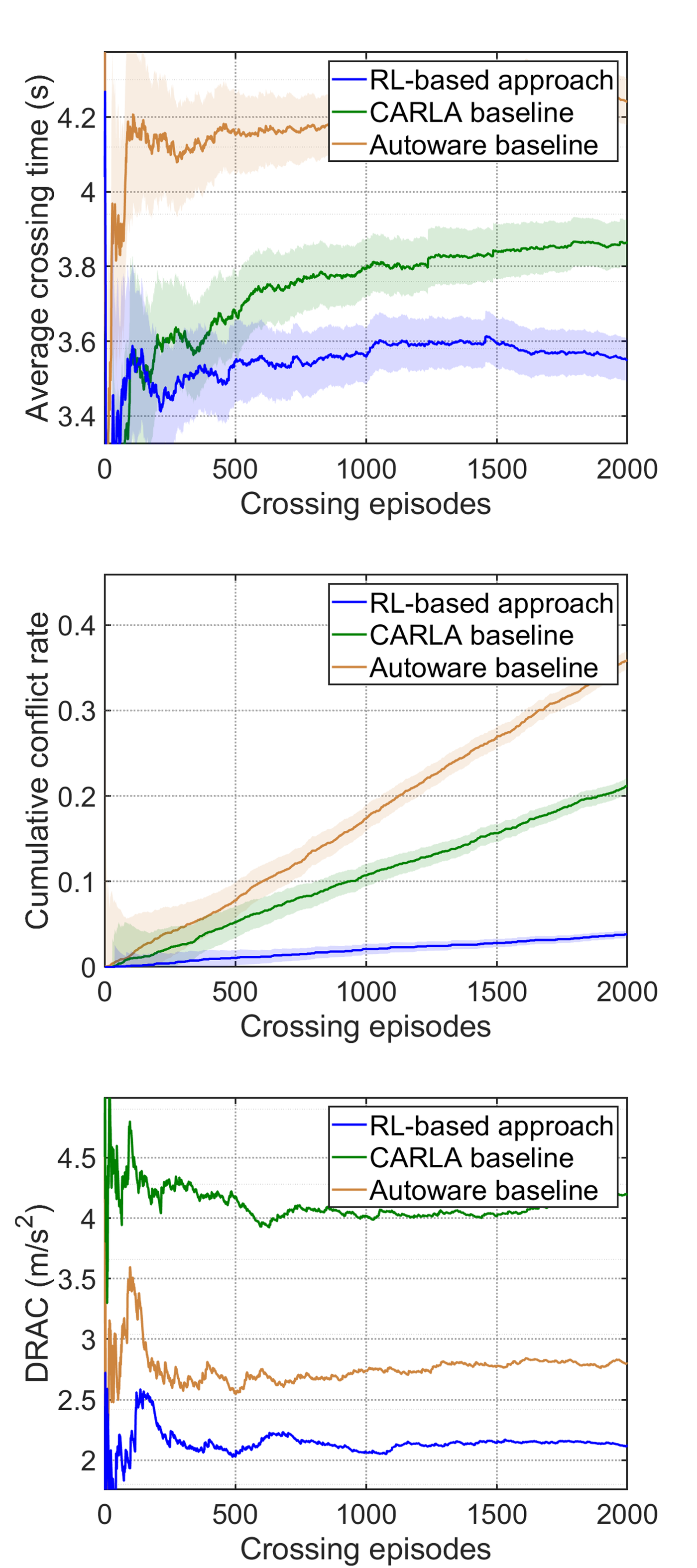}
        \centerline{(c) Tokushima}
    \end{minipage}
    \caption{Performance comparison across three real-world intersection scenarios in terms of efficiency (top), safety (middle), and comfort (bottom)}
    \label{fig:sim}
\end{figure*}

To quantify the contribution of each module and compare with representative baselines, five comparison groups are analyzed: (1) an online-only MAPPO baseline; (2) a model with offline pre-training but without self-attention; (3) a model trained with a mixed replay buffer rather than direction-specific buffers; (4) Multi-Agent Deep Deterministic Policy Gradient (MADDPG) \cite{lowe2020multiagentactorcriticmixedcooperativecompetitive}; and (5) Social Learning Policy Optimization (SoLPO)\cite{11036678}. All methods are compared in terms of average reward, success rate, and convergence speed.

Fig.~\ref{res} (b) and (c) illustrate the convergence of the proposed approach and the baselines. The complete approach converges within approximately 250 episodes, whereas the online-only baseline requires more than 800 episodes. Removing the self-attention module slows convergence to around 500 episodes, confirming that attention is critical for modeling complex multi-agent interactions and maintaining coordination stability. The mixed replay buffer variant converges after roughly 600 episodes and achieves lower final performance, indicating that direction-specific buffers effectively mitigate policy conflict among heterogeneous maneuvers and reduce class imbalance during pre-training. Compared with the other representative baselines, both methods converge more slowly than the proposed approach without offline pre-training. MADDPG also shows inferior final performance, while SoLPO reaches a similar reward level but remains less stable in terms of success rate. These results demonstrate that both the self-attention mechanism and the tailored offline data organization are key to improving sample efficiency and achieving robust multi-agent coordination. Overall, the proposed two-stage training framework significantly improves policy generalization and stability in smart intersection scenarios.

\subsection{Multi-Intersection Simulation Evaluation}

To further evaluate the generalization and scalability of the proposed system, three representative real-world intersections in Japan, located in  Fukuoka, Hyogo, and Tokushima, were selected based on publicly traffic accident statistics \cite{sonpo2024kousaten,mlit2024shikoku,rkb2022fukuoka}, as illustrated in Fig.~\ref{m}. These sites encompass diverse urban traffic configurations, ranging from a simple T-junction with side merging, a ramp merging into a main road, and a multi-lane intersection with converging through traffic. Such diversity enables a comprehensive assessment of the proposed system under heterogeneous and realistic traffic conditions.

The Fukuoka site is a signalized intersection with a geometric layout, where three through lanes merge into two downstream of the intersection. It is suitable for evaluating multi-lane merging coordination. The Hyogo site is an unsignalized ramp-merging intersection, where vehicles must enter a high speed road within a short acceleration lane, placing high demands on cooperative yielding and merging efficiency. The Tokushima site is an unsignalized T-junction with a side-merging configuration, where vehicles from the minor road must find gaps in the main-road traffic flow to merge, leading to typical lateral interaction conflicts. These intersections differ in right-of-way rules, interaction complexity, and conflict patterns, providing a comprehensive basis for evaluating the robustness and generalization of the proposed system under diverse traffic conditions.

\begin{figure*}[!t]
    \centerline{\includegraphics[width=0.9\textwidth]{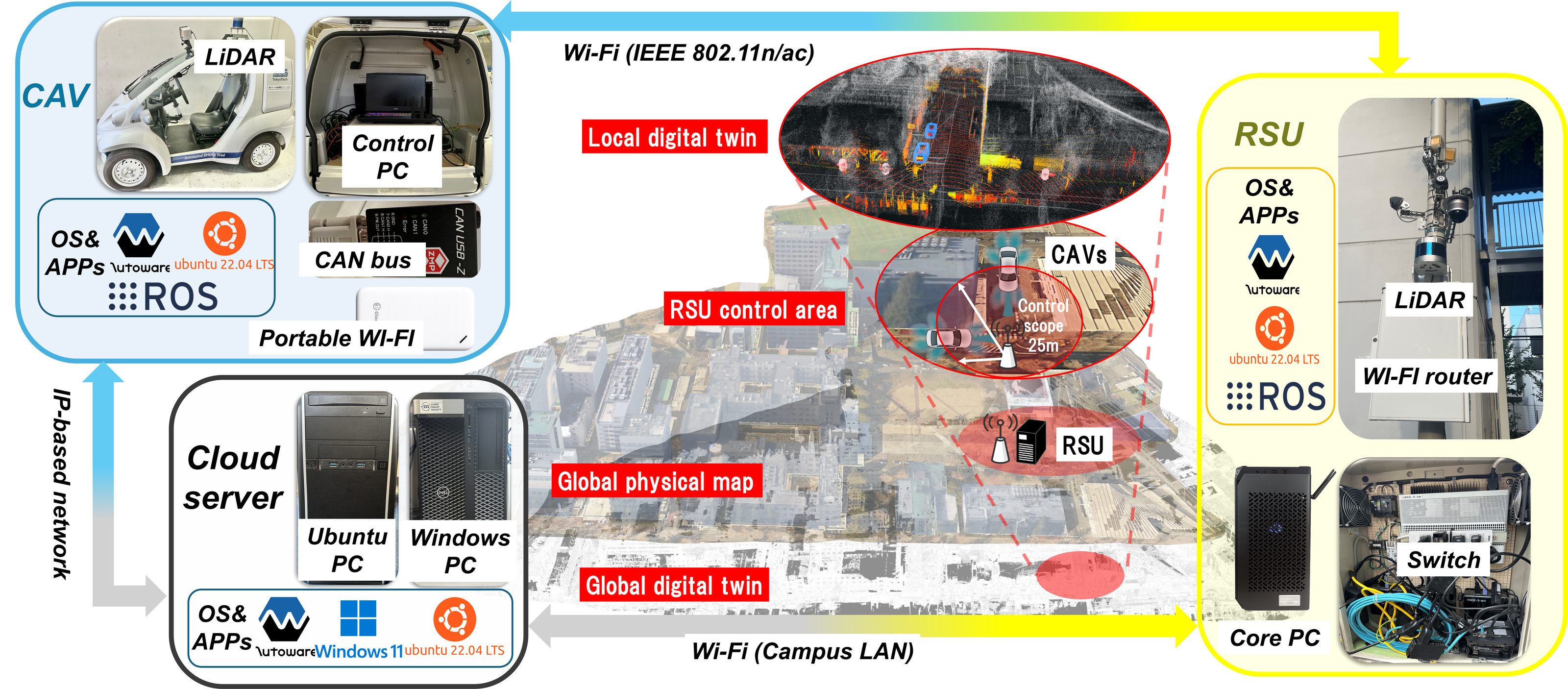}}
    \caption{System architecture of the PoC for cooperative driving platform}
    \label{fig: PoC}
\end{figure*}

Using Google Street View data, we constructed high-fidelity digital simulation environments for the three selected intersections and deployed closed-loop traffic scenarios in CARLA. Each scenario maintains a constant traffic flow of 20 vehicles, including 6 CAVs and 14 HDVs, with randomly generated pedestrians to preserve complexity. A fixed delay of 50 ms is applied in simulation to approximate the combined transmission and decision delay observed in the real-world system, while perception and execution latencies are analyzed separately in the real-world implementation. The performance of 2000 CAV passages through the intersection is recorded to examine the stability and generalization of the proposed system under closed-loop traffic.

The performance of the proposed centralized control system was compared with two baseline methods, a distributed vehicle-level control module based on Autoware and a rule-based local control module implemented in CARLA. Performance is evaluated in terms of traffic efficiency, safety, and comfort. Average travel time is used to measure intersection efficiency, the cumulative conflict rate reflects safety and potential collision risks, and the average deceleration rate to avoid a crash (DRAC) metric quantifies deceleration smoothness and passenger comfort. Each reported result is obtained from repeated long-horizon closed-loop evaluations over 2000 CAV passages, so as to reflect the performance trends of the proposed framework under realistic multi-agent traffic interactions. The cumulative conflict rate at the $n$-th crossing episode is defined as
\begin{equation}
R_{\mathrm{conf}}(n)=\frac{N_{\mathrm{conf}}^{\mathrm{total}}(n)}{n},
\end{equation}
where $N_{\mathrm{conf}}^{\mathrm{total}}(n)$ is the total number of conflict events observed up to the $n$-th completed CAV passage.

The results across the three intersections are presented in Fig.~\ref{fig:sim}. The proposed centralized control system outperforms baseline approaches in terms of traffic efficiency, safety, and comfort. As the number of CAV passages increases, our proposed system converges rapidly and maintains the lowest average travel time. In terms of safety, it achieves the lowest conflict rate. In terms of comfort, it shows the lowest mean and variance of DRAC, indicating smooth cooperative merging with gentle deceleration while maintaining safety.

In contrast, the distributed Autoware baseline lacks global traffic perception, as each vehicle makes decisions independently based on local perception and rule-based logic. This often results in merging conflicts and repeated mutual yielding near conflict points, which increases travel time and leads to the accumulation of interaction conflicts. Its conservative control policy and inherent safety margins yield relatively stable DRAC performance. The CARLA baseline, although rule-based and focused on basic safety constraints, lacks high-level interactive decision-making and negotiation mechanisms. Consequently, it cannot solve conflicts among CAVs as traffic density increases, causing a continuous growth in safety risk. Moreover, because its controller prioritizes vehicle dynamics realism over ride comfort, sudden braking and unnecessary stopping frequently occur in complex scenarios, leading to greater DRAC fluctuations and reduced comfort.

Overall, the proposed approach maintains superior and stable performance in closed-loop traffic flow and generalizes effectively to real-world intersections with distinct structural characteristics, achieving comprehensive improvement in efficiency, safety, and comfort.

%% file: 6_PoCset.tex
\begin{figure}[t]
    \centering
    \includegraphics[width=0.49\textwidth]{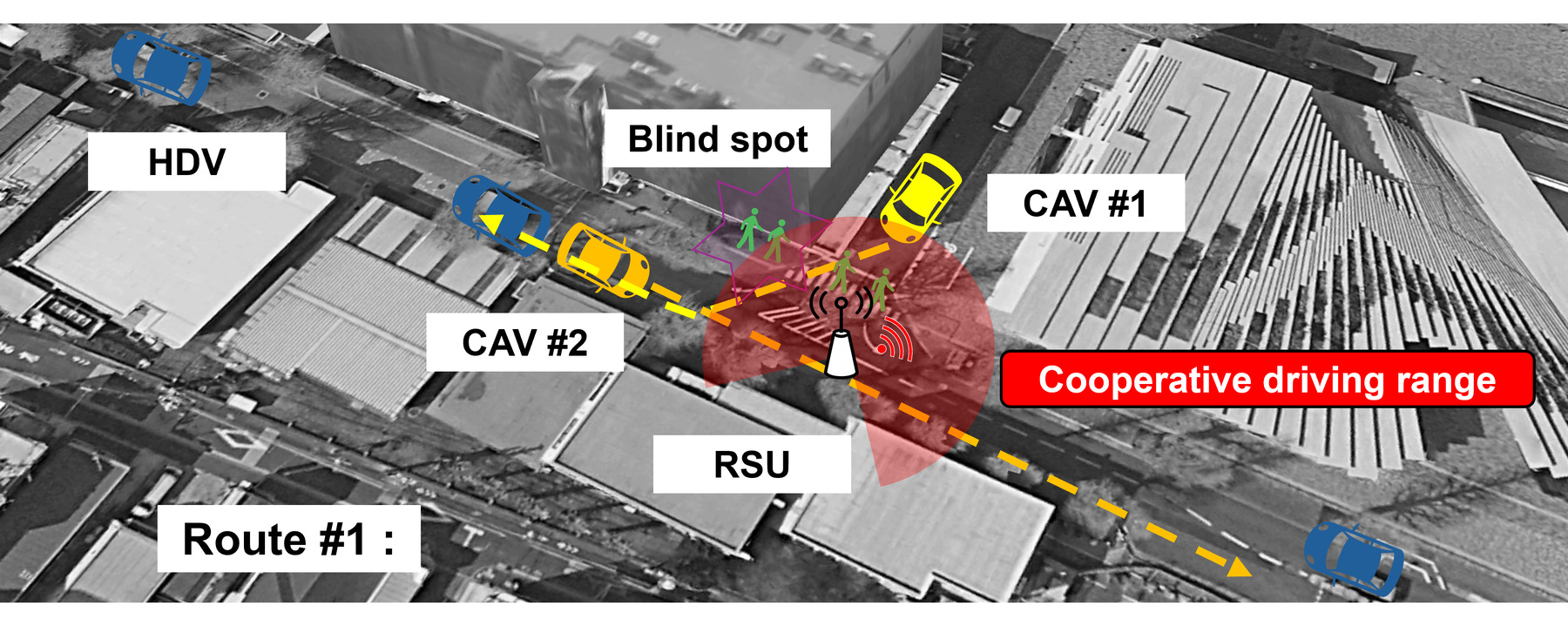}
    \centerline{(a)}
    
    \vspace{2mm}
    
    \includegraphics[width=0.49\textwidth]{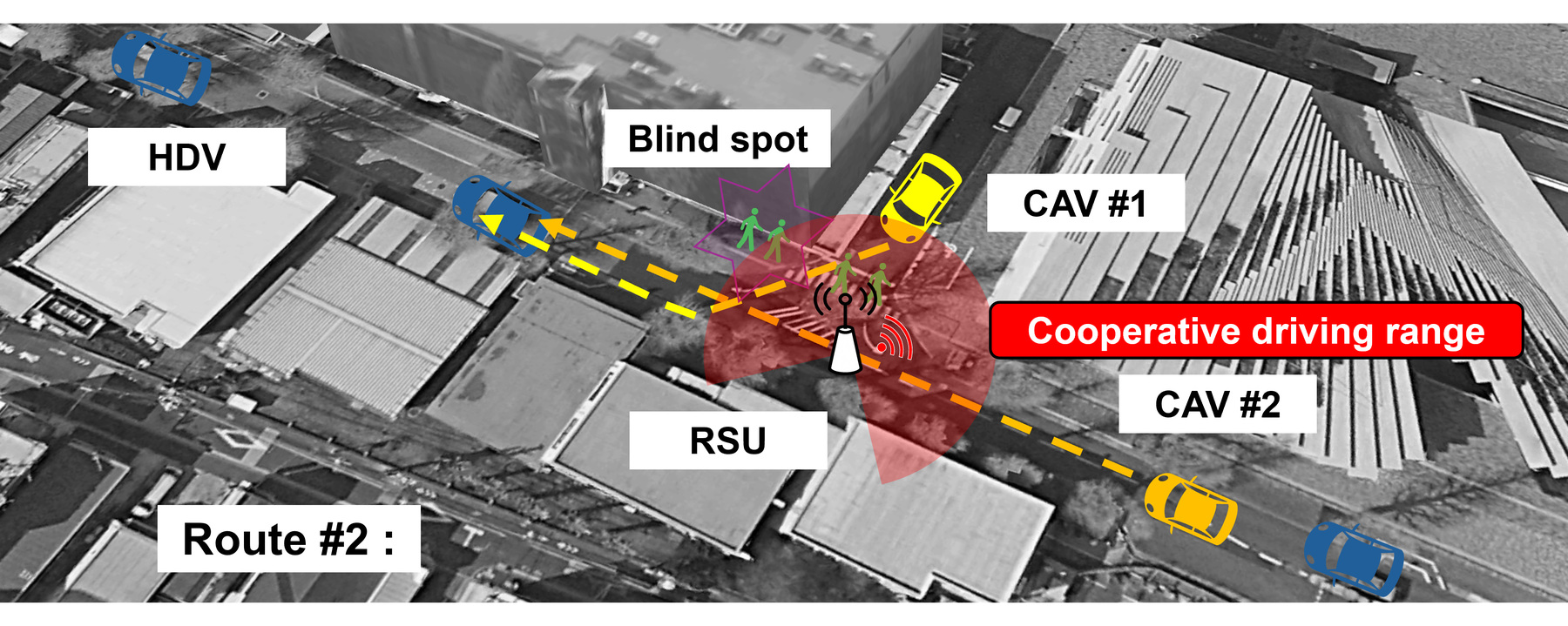}
    \centerline{(b)}
    \caption{Experimental routes and cooperative driving range for the PoC validation: (a) crossing scenario, where CAV~\#1 travels from north to south, and CAV~\#2 travels from west to east under blind spot area; (b) merging scenario, where CAV~\#1 travels from north to south preparing to merge into main road, and CAV~\#2 travels from east to west}
    \label{fig: route}
\end{figure}

This section introduces the PoC platform built to evaluate the proposed V2I2V system in the real world. It includes the hardware and software architecture, communications design, and PoC setup. Fig.~\ref{fig: PoC} shows the overall system architecture and communications links, while Fig.~\ref{fig: route} illustrates the test routes designed for PoC validation. The main hardware and software configurations of the PoC platform are summarized in Table~\ref{tab:hw_specs}.

\begin{table*}[t]
\centering
\caption{Hardware Configuration of the RSU-Centric Cooperative Driving Platform}
\label{tab:hw_specs}
\begin{tabular}{c|c|p{10.5cm}}
\hline
\textbf{Type} & \textbf{Device} & \textbf{Specifications} \\ \hline

\multirow{2}{*}{\textbf{CAV}}
& COMS (CAV~\#1) & Drive-by-wire vehicle equipped with a 32-layer LiDAR (200~m range, $\pm$3~cm accuracy), onboard control laptop (Intel~Core~i7-10750H~CPU, NVIDIA~GeForce~RTX~3060~GPU, Ubuntu~16.04~LTS, ROS~1~Kinetic, Autoware~AI), CANBUS module, and portable Wi-Fi router. \\ \cline{2-3}
& Virtual CAV (CAV~\#2) & Digital vehicle instantiated in the Local DT. \\ \hline

\multirow{2}{*}{\textbf{Sensors}}
& CAV LiDAR & 32-layer LiDAR (200~m range, $\pm$3~cm accuracy). \\ \cline{2-3}
& RSU LiDAR & 64-layer LiDAR (200~m range, $\pm$3~cm accuracy). \\ \hline

\multirow{2}{*}{\textbf{Edge\&Cloud Servers}}
& RSU Core PC & Industrial PC with AMD~Ryzen~5~9600~CPU and NVIDIA~RTX~4070~Ti~Super~GPU; Ubuntu~22.04~LTS, ROS~2~Humble, Autoware~Universe. \\ \cline{2-3}
& Cloud Workstation & Intel~Core~i9~CPU, NVIDIA~RTX~3090~GPU; Ubuntu~22.04~LTS and Windows~11 dual system. \\ \hline

\multirow{3}{*}{\textbf{Network Devices}}
& Portable Wi-Fi (CAV) & IEEE~802.11n/ac router for V2I/V2C links. \\ \cline{2-3}
& Wi-Fi AP (RSU) & IEEE~802.11n/ac access point for V2I and I2C connections via campus LAN. \\ \cline{2-3}
& Ethernet Switch & Gigabit Ethernet switch (1000~Mbps, 8~ports) for RSU-side sensor and compute connections. \\ \hline

\end{tabular}
\end{table*}

\subsection{Hardware Deployment and Communication Design}

The RSU-centric V2I2V cooperative driving system was built on the Institute of Science Tokyo campus, as illustrated in Fig.~\ref{fig: PoC}. The system consists of three main components: CAVs, the RSU server, and the cloud server. Each CAV is equipped with a 32-layer LiDAR, an onboard industrial computer, a CANBUS interface, and a portable Wi-Fi router. The RSU is installed near the intersection and integrates a 64-layer LiDAR with a high-performance industrial PC featuring an AMD Ryzen 5 9600 CPU and an NVIDIA RTX 4070 Ti Super GPU for real-time perception and cooperative driving decision-making. The cloud server maintains the global DT for global data collection and high-level monitoring. The proposed RSU-centric architecture concentrates the additional sensing and computing resources at the intersection as shared infrastructure, rather than duplicating them on every CAV, which further reduces the overall system deployment cost.

For communications, a unified Wi-Fi network (IEEE 802.11n/ac) is adopted as the medium for V2I2V interactions among CAVs, the RSU, and the cloud. Each CAV transmits real-time perception data and receives control commands through a portable Wi-Fi router connected to the RSU, while the RSU communicates with the cloud for global data synchronization. Although Wi-Fi does not achieve the ultra-low latency and reliability typically associated with dedicated short-range communications (DSRC), its low deployment cost, portability, and widespread availability make it a practical medium for PoC validation. The reported PoC communication and decision latencies are measured after the Wi-Fi link is established. The initial scanning, association, and setup delay is treated as part of the access and control handover process.

Unlike previous studies that rely on multiple physical CAVs, this paper adopts a mixed-reality validation approach that integrates one physical and one virtual CAV. The physical vehicle (CAV~\#1) performs real-world perception and transmits sensor data to the RSU, which computes control commands within the local DT and sends them back for closed-loop execution. The virtual vehicle (CAV~\#2) is instantiated in the local DT as a digitally twinned counterpart with identical control interfaces. The RSU coordinates both CAVs and sends control signals through the same decision-making pipeline, ensuring that the virtual CAV mirrors the operational logic of the physical one. This design not only avoids the cost and safety risks of deploying two real vehicles in the confined campus environment with dense pedestrian traffic but also preserves experimental fidelity while improving repeatability and scalability for cooperative driving validation.

\begin{table*}[t]
\centering
\caption{Quantitative Results of the Proposed and Baseline Approaches Benchmarked against 3GPP V2X Requirements}
\label{tab:poc_metrics}
\begin{tabular}{c|c|c|c|c}
\hline
\textbf{Metric} & \textbf{3GPP Requirement} & \textbf{Baseline} & \textbf{Proposed Method} & \textbf{Improvement} \\ \hline
\textbf{$T_{\mathrm{dec}}$ (ms)} & $500$–$1000$ & $74 \pm 8$ & $\mathbf{42 \pm 6}$ & $43.2\%$ \\ \hline
\textbf{Stopping Distance to Pedestrian (m)} & N/A & $3.2 \pm 0.3$ & $\mathbf{8.5 \pm 1.1}$ & $165.6\%$ \\ \hline
\textbf{Stop Waiting Time (s)} & N/A & $6.8 \pm 0.6$ & $\mathbf{4.5 \pm 0.4}$ & $33.8\%$ \\ \hline
\textbf{Total Passing Time (s)} & N/A & $15.2 \pm 1.0$ & $\mathbf{12.1 \pm 0.8}$ & $20.4\%$ \\ \hline
\end{tabular}
\end{table*}

\begin{figure}[t]
    \centering
    \includegraphics[width=0.49\textwidth]{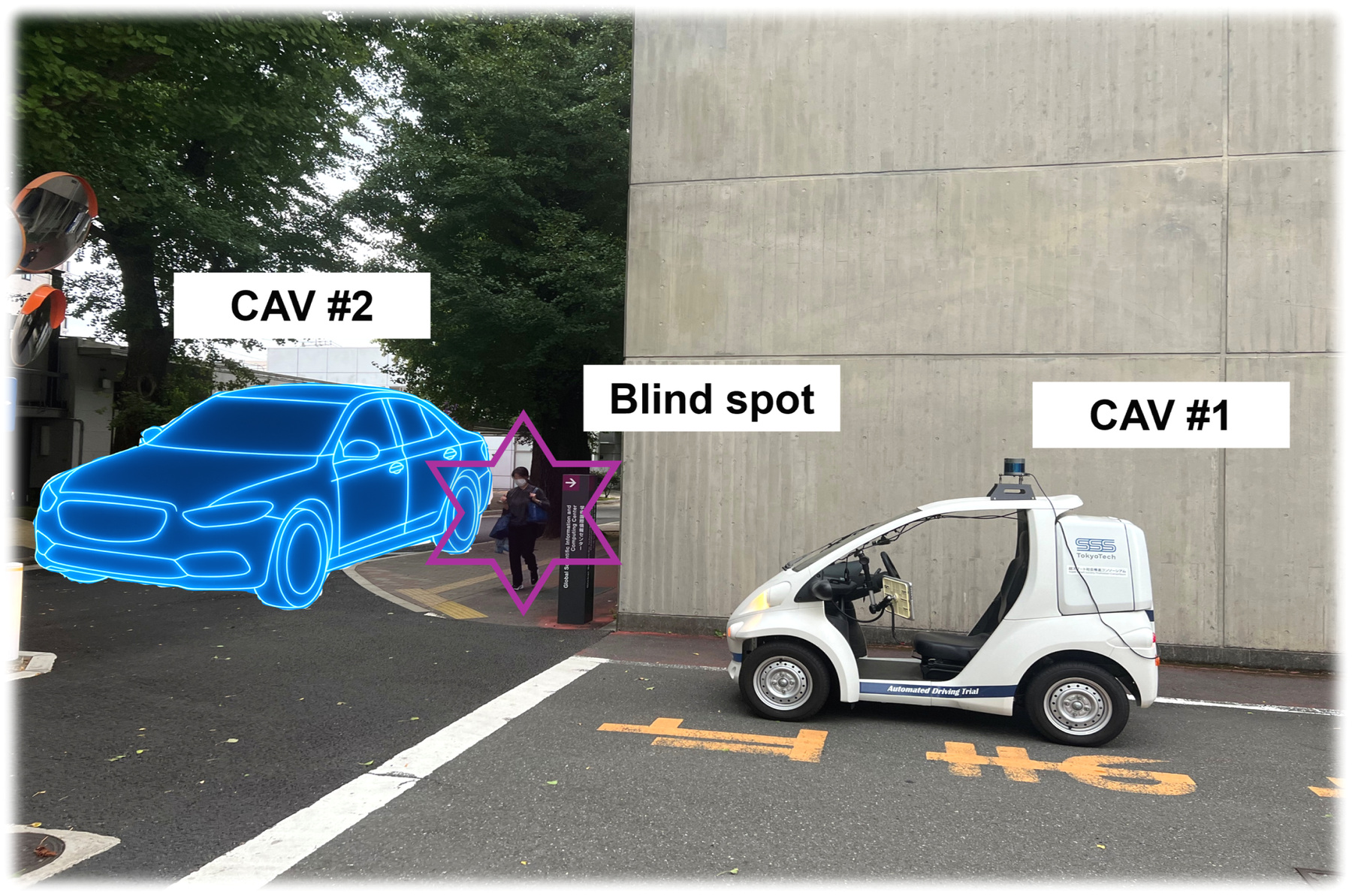}
    
    \vspace{2mm}
    
    \includegraphics[width=0.49\textwidth]{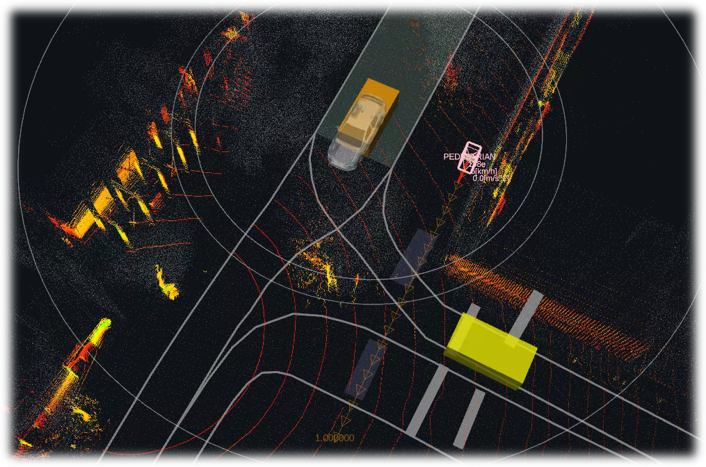}
    \caption{Mixed-reality blind spot scenario for V2I2V cooperative driving}
    \label{fig: tt}
\end{figure}

For intersection cooperative control, an RSU service area with a 25\,m radius is defined. When CAVs enter this area, control authority is transferred from the onboard controller to the RSU after the cooperative communication link is established and the remote control conditions are satisfied. Before this handover is completed, the vehicle Autoware local autonomy stack remains responsible for basic driving and safety maintenance, so the vehicle is not left uncontrolled during the initial access process. Such access and control handover procedures can be supported by heterogeneous network architectures, as investigated in our previous work~\cite{10185532}. The RSU then computes coordinated trajectories in the local DT and guides the vehicles through the intersection. After exiting the service area, the CAVs resume autonomous driving. This mechanism demonstrates the feasibility of RSU-centric V2I2V cooperative control and enhances safety at smart intersections.

\begin{figure}[t]
    \centerline{\includegraphics[width=0.49\textwidth]{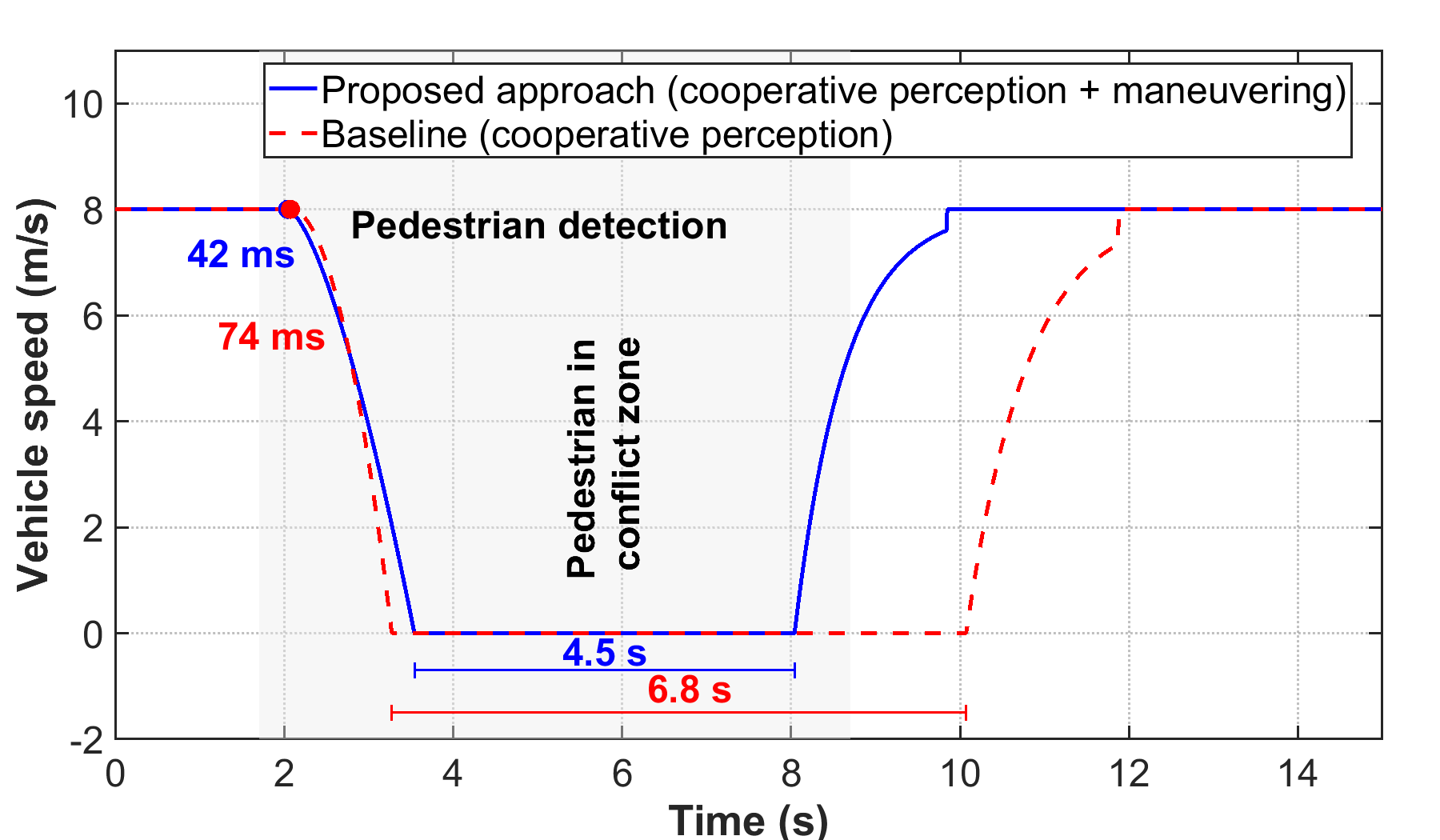}}
    \caption{Comparison of responses between the proposed approach and the baseline under cooperative perception, where cooperative maneuvering is enabled only in the proposed approach}
    \label{fig: gg}
\end{figure}

\subsection{Experimental Scenarios and Test Setup}

To validate the proposed V2I2V system, intersection experiments were conducted on two test routes within the same traffic environment containing both HDVs and pedestrians, as shown in Fig.~\ref{fig: route}. The two routes feature different approach directions, enabling comprehensive assessments of intersection negotiation performance.

In both routes, the RSU continuously monitors surrounding road users through the local DT and sends cooperative control signals to the CAVs to ensure safe and efficient intersection, especially under blind spots and potential conflict scenarios. During the experiments, CAV trajectories, RSU processing delays, and end-to-end (E2E) latency are recorded for quantitative performance evaluation.

To clarify the E2E latency composition, we decompose it as
\begin{equation}
T_{\mathrm{E2E}} = T_{\mathrm{perc}} + T_{\mathrm{trans}} + T_{\mathrm{dec}} + T_{\mathrm{exec}},
\end{equation}
where $T_{\mathrm{perc}}$ is the RSU perception latency from sensing to structured state generation, $T_{\mathrm{trans}}$ is the one-way communication latency between the RSU and the CAV, $T_{\mathrm{dec}}$ is the inference latency of the cooperative decision model, and $T_{\mathrm{exec}}$ is the latency from the reception of a control command to the first observable physical response. In our prototype, $T_{\mathrm{perc}} \approx 28.8~\mathrm{ms}$, $T_{\mathrm{trans}} \approx 8.5~\mathrm{ms}$, and $T_{\mathrm{exec}} \approx 100.0~\mathrm{ms}$. The inference latency $T_{\mathrm{dec}}$ is further reflected in the baseline comparison reported in Table~\ref{tab:poc_metrics} and illustrated in Fig.~\ref{fig: gg}.

%% file: 7_evaluation.tex
To validate the proposed V2I2V cooperative driving system in the real world, a PoC trial was conducted on a university campus. The test scenario is shown in Fig.~\ref{fig: route}, which features a smart intersection similar to the training environment used in simulation. The system adopts a centralized perception and distributed actuation architecture. The PoC trial is intended to validate the real-world feasibility of the proposed system, including the E2E loop of perception, communication, decision-making, and execution, while the multi-intersection simulations complement this validation by evaluating scalability and generalization under more complex traffic conditions. On each CAV, ROS nodes periodically report sensing data and vehicle pose and velocity via HTTP to the RSU. The RSU is equipped with LiDAR and sensor fusion modules, and ROS edge computing nodes receive perception data, execute the centralized decision-making modules, generate longitudinal and lateral control signals, and send them back to the vehicles through HTTP for execution\cite{10185532}. This design achieves centralized perception at the RSU, enabling coordinated interactions among multiple CAVs at the intersection with low latency.

To provide a comparison against cooperative control, a non-cooperative maneuvering baseline was established. In this setting, the RSU is limited to perception and object detection, and only forwards environmental information to the vehicles. Each CAV then performs local trajectory planning and control independently using its onboard Autoware system, without any coordination or shared decision-making. As a result, while cooperative perception is available in the baseline, maneuvering decisions are made independently at the vehicle level.

Fig.~\ref{fig: tt} illustrates the blind-spot scenario used in the PoC experiment, showing both the physical intersection and its digital replica in the local DT. Fig.~\ref{fig: gg} compares the speed–time profiles of the proposed approach and the baseline under the same cooperative perception conditions. When pedestrians enter the blind area, both approaches are able to detect the pedestrian and initiate braking. In the baseline, although cooperative perception is available, each CAV performs self-driving maneuvering independently based on its onboard Autoware system. The lack of coordinated maneuvering leads to delayed and more abrupt braking behaviors, resulting in a shorter stopping distance, a longer stop duration, and a slower restart, which increases the overall traversal time. In contrast, the proposed approach enables cooperative maneuvering among multiple CAVs based on decision-making on RSU. By jointly regulating vehicle responses, approaching vehicles decelerate earlier and more smoothly, resulting in larger stopping distances and improved safety and ride comfort.

Table~\ref{tab:poc_metrics} summarizes the quantitative results obtained from repeated PoC trials. The proposed approach achieves consistent improvements in efficiency, safety, and ride comfort compared with the baseline. Specifically, the minimum stopping distance increases by 165.6\%, allowing earlier and smoother deceleration and reducing collision risk. From a system perspective, the proposed approach also reduces average decision latency by approximately 43\%, while remaining well below the V2X latency threshold specified by 3GPP~\cite{3gpp_v2x_spec}. Stop duration and overall traversal time are reduced, leading to improved intersection throughput.

Overall, the PoC results demonstrate that the proposed V2I2V cooperative driving system achieves substantial improvements in multiple CAVs coordination, traffic efficiency, and operational safety in real-world traffic environments. The SAE J2735 and J2945 \cite{sae_j2735_202409,sae_j2945_1_202004} are closely related to this work, since they provide a standardized framework for V2X message exchange and communication performance requirements. In our system, such standards offer an important communication and interoperability foundation, while the main contribution of this paper lies in the RSU-centric cooperative perception, DT-based decision-making, and intersection coordination built on top of that foundation.

%% file: 8_conclusion.tex
This paper introduces a DT-based V2I2V cooperative driving system to improve safety and efficiency at smart intersections. By integrating global and local DTs with RSU-centric decision-making, the proposed system enables robust and adaptive multi-vehicle coordination and generalizes across diverse intersection layouts. A real-world PoC trial verified the feasibility of the V2I2V architecture, where centralized perception and RSU decision-making significantly reduced end-to-end latency and improved pedestrian safety. Furthermore, multi-intersection simulations across three representative real-world sites confirmed the scalability and generalization capability of the system, maintaining low conflict rates under communications latency and achieving high traffic throughput at high CAV penetration levels. Together, the real-world PoC and multi-intersection simulation results demonstrate the potential of the proposed RSU-centric V2I2V cooperative driving system for next-generation ITS, with the former validating real-world feasibility and the latter confirming scalability and generalization under more complex traffic conditions. Future work will focus on validation with multiple CAVs in more complex open-road environments, including denser heterogeneous traffic interactions and diverse environmental conditions.